\documentclass[preprint2]{proto}
\usepackage{times}

\newcommand{\refs}{\par\noindent\hangindent=1pc\hangafter=1}
\def\msun{\rm M_{\odot}}

\newbox\grsign \setbox\grsign=\hbox{$>$} \newdimen\grdimen \grdimen=\ht\grsign
\newbox\simlessbox \newbox\simgreatbox
\setbox\simgreatbox=\hbox{\raise.5ex\hbox{$>$}\llap
     {\lower.5ex\hbox{$\sim$}}}\ht1=\grdimen\dp1=0pt
\setbox\simlessbox=\hbox{\raise.5ex\hbox{$<$}\llap
     {\lower.5ex\hbox{$\sim$}}}\ht2=\grdimen\dp2=0pt
\def\simgreat{\mathrel{\copy\simgreatbox}}
\def\simless{\mathrel{\copy\simlessbox}}

\def\GMCM2{\rm gm\,cm^{-2}}

\def\apj{\it Astrophys. J.\rm}
\def\apjs{\it Astrophys. J. Suppl.\rm}
\def\aap{\it Astron. Astrophys.\rm}
\def\aaps{\it Astron. Astrophys. Suppl.\rm}
\def\aj{\it Astron. J.\rm}
\def\mnras{\it Mon. Not. R. Astron. Soc.\rm}
\def\pasp{\it Publ. Astron. Soc. Pac.\rm}
\def\pasj{\it Publ. Astron. Soc. Japan\rm}

\def\science{\it Science\rm}

\def\araa{\it Ann. Rev. Astron. Astrophys.\rm}

\begin{document}

\title{\textbf{\LARGE The Low-mass Populations in OB Associations}}

\author {\textbf{\large C\'esar Brice\~no}}
\affil{\small\em Centro de Investigaciones de Astronom{\'\i}a}
\author {\textbf{\large Thomas Preibisch}}
\affil{\small\em Max-Planck-Institut f\"ur Radioastronomie}
\author {\textbf{\large William H. Sherry}}
\affil{\small\em  National Optical Astronomy Observatory}
\author {\textbf{\large Eric E. Mamajek}}
\affil{\small\em Harvard-Smithsonian Center for Astrophysics}
\author {\textbf{\large Robert D. Mathieu}}
\affil{\small\em University of Wisconsin - Madison}
\author {\textbf{\large Frederick M. Walter}}
\affil{\small\em Stony Brook University}
\author {\textbf{\large Hans Zinnecker}}
\affil{\small\em Astrophysikalisches Institut Potsdam}

\begin{abstract}
\baselineskip = 11pt
\leftskip = 0.65in
\rightskip = 0.65in
\parindent=1pc 
{\small 
Low-mass stars ($\rm 0.1 \la M \la 1~ \msun$)
in OB associations are key to addressing some of
the most fundamental problems in star formation.
The low-mass stellar populations of OB associations
provide a snapshot of the fossil 
star-formation record of giant molecular cloud complexes.
Large scale surveys have identified hundreds
of members of nearby OB associations, and revealed that low-mass stars
exist wherever high-mass stars have recently formed. The spatial
distribution of low-mass members of OB associations demonstrate the existence 
of significant substructure ("subgroups"). 
This "discretized" sequence of stellar groups is consistent with an origin
in short-lived parent molecular clouds within a Giant Molecular Cloud
Complex.  The low-mass population in each subgroup within an OB
association exhibits little evidence for significant age spreads
on time scales of $\sim 10$ Myr or greater, in
agreement with a scenario of rapid star formation and cloud
dissipation.
The Initial Mass Function (IMF) of the stellar populations in OB associations
in the mass range $\rm 0.1 \la M \la 1~\msun$ is largely
consistent with the field IMF, and
most low-mass pre-main sequence stars in the solar vicinity are in
OB associations. These findings agree with early suggestions 
that {\it the majority of stars in the Galaxy were born in OB associations}.
The most recent work further suggests that a significant fraction of
the stellar population may have their origin in the more spread
out regions of OB associations, instead of all being born in dense clusters.
Ground-based and space-based (Spitzer
Space Telescope) infrared studies have provided robust evidence that
primordial accretion disks around low-mass stars dissipate on
timescales of a few Myr.  However, on close inspection
there appears to be great variance in the
disk dissipation timescales for stars of a given mass in OB
associations. While some stars appear to lack disks at $\sim$1 Myr,
a few appear to retain accretion disks up to ages of
$\sim$10-20 Myr.
 \\~\\~\\~}
\end{abstract}  

\section{\textbf{INTRODUCTION}}

Most star formation in normal galaxies occurs in the cores of the
largest dark clouds in spiral arms, known as Giant Molecular Clouds (GMCs).
A GMC may give rise to one or more star complexes known as OB associations,
first defined and recognized by {\it Ambartsumian} (1947) as young
expanding stellar systems of blue luminous stars. These generally
include groups of T Tauri stars or T associations
({\it Kholopov,} 1959; {\it Herbig,} 1962; {\it Strom et al.,} 1975) as well as clusters,
some containing massive (M $\ga 10~\msun$) stars, but all teeming with
solar-like and lower mass stars.

Though we now recognize OB associations as the prime sites for star formation
in our Galaxy, much of our knowledge of star formation is based on
studies of low-mass (M $\la 1~\msun$) pre-main sequence (PMS) stars located
in nearby T associations, 
like the $\sim 1-2$ Myr old Taurus, Lupus and Chamaeleon star forming regions.
The view of star formation conveyed by these observations
is probably biased to the particular physical conditions 
found in these young, quiescent regions.
In contrast, the various OB associations in the solar vicinity
are in a variety of evolutionary stages and environments,
some containing very young objects (ages $\la 1$~Myr)
still embedded in their natal gas (e.g., Orion A and B clouds, Cep OB2), 
others in the process of dispersing their parent clouds, like
$\lambda$ Ori and Carina,
while others harbor more evolved populations, several Myr old, 
which have long since dissipated their progenitor clouds 
(like Scorpius-Centaurus and Orion OB 1a). 
The low-mass populations in these differing regions
are key to investigating fundamental issues in the formation and
early evolution of stars and planetary systems:

\noindent
1) {\it Slow vs. rapid protostellar cloud collapse and
molecular cloud lifetimes.}
 In the old model of star formation (see {\it Shu et al.}, 1987)
protostellar clouds
contract slowly until ambipolar diffusion removes enough
magnetic flux for dynamical (inside-out) collapse to set in. 
It was expected that the
diffusion timescale of $\sim$~10 Myr should produce
a similar age spread in the resulting populations of stars,
consistent with the $\la$ 40 Myr early estimates of molecular
cloud lifetimes (see discussion in {\it Elmegreen}, 1990). Such age
spreads should be readily apparent in color-magnitude or H-R diagrams 
for masses $\la 1\>\msun$.
However, the lack of even $\sim$ 10 Myr old, low-mass stars in and near
molecular clouds challenged this paradigm,
suggesting that star formation
proceeds much more rapidly than previously thought, even over
regions as large as 10 pc in size ({\it Ballesteros-Paredes et al.}, 1999),
and therefore that cloud lifetimes over the same scales could
be much shorter than 40 Myr ({\it Hartmann et al.}, 1991).

\noindent
2) {\it The shape of the IMF.}
Whether OB associations have low mass populations according
to the field IMF, or if their IMF is truncated 
is still a debated issue.
There have been many claims for IMF cutoffs in
high mass star forming regions 
(see e.g., {\it Slawson and Landstreet}, 1992; {\it Leitherer}, 1998; 
{\it Smith et al.}, 2001; {\it Stolte et al.}, 2005). 
However, several well investigated massive star forming regions show
{\em no} evidence for an IMF cutoff
(see {\it Brandl et al.}, 1999 and {\it Brandner et al.}, 2001 
for the cases of NGC~3603 and 30~Dor, respectively),
and notorious difficulties in IMF determinations of distant regions
may easily lead to wrong conclusions about IMF variations
(e.g., {\it Zinnecker et al.}, 1993; {\it Selman and Melnick}, 2005).
An empirical proof of a field-like IMF, rather than a truncated IMF, has
important consequences not only for star formation models but also for
scenarios of distant starburst regions; 
e.g., since most of the stellar mass is then in low-mass stars, this 
limits the amount of material which is enriched in metals via 
nucleosynthesis in massive stars and which is then injected back 
into the interstellar medium by the winds and supernovae of the massive stars.

\noindent
3) {\it Bound vs. unbound clusters.}
While many young stars are born in groups and clusters, most disperse
rapidly; few clusters remain bound over timescales $> 10$~Myr.
The conditions under which bound clusters
are produced are not clear.  Studies of older, widely-spread low-mass
stars around young clusters might show a time sequence of cluster formation,
and observations of older, spreading
groups would yield insight into how and why clusters disperse.

\noindent
4) {\it Slow vs. rapid disk evolution.}
Early studies of near-infrared dust emission from low-mass 
young stars suggested that most stars lose their optically thick disks
over periods of $\sim 10$~Myr,
(e.g., {\it Strom et al.}, 1993), similar to the timescale
suggested for planet formation ({\it Podosek and Cassen}, 1994).
However, there is also evidence for faster evolution in some cases;
for example, half of all $\sim 1$ Myr-old stars in Taurus have strongly reduced
or absent disk emission ({\it Beckwith et al.}, 1990).
The most recent observations of IR emission from low-mass PMS stars
in nearby OB associations like Orion, suggest that the 
timescales for the dissipation of the inner disks can vary
even in coeval populations at young ages
({\it Muzerolle et al.}, 2005).

\noindent
5) {\it Triggered vs. independent star formation.}
Although it is likely that star formation in one region can ``trigger''
more star formation later in neighboring areas, and there is evidence for this
from studies of the massive stars in OB populations
(e.g., {\it Brown}, 1996), proof
of causality and precise time sequences are difficult to obtain without
studying the associated lower mass populations.
In the past, studies of
the massive O and B stars have been used to investigate sequential star formation
and triggering on large scales (e.g., {\it Blaauw}, 1964, 1991 and references therein).
However, OB stars are formed essentially on the main sequence
(e.g., {\it Palla and Stahler} 1992, 1993) and evolve off the main sequence on a
timescale of order 10 Myr (depending upon mass
and amount of convective overshoot), thus they are not useful tracers of
star-forming histories on timescales of several Myr, while
young low-mass stars are.
Moreover, we cannot investigate cluster structure and dispersal or
disk evolution without studying low-mass stars.  Many young
individual clusters have been studied at both optical and infrared
wavelengths (c.f. {\it Lada and Lada} 2003), but these only represent the highest-density
regions, and do not address older and/or more widely dispersed populations.
In contrast to their high mass counterparts, low-mass stars 
offer distinct advantages to address the aforementioned issues. 
They are simply vastly more numerous than O, B, and A stars,
allowing statistical studies not possible with the few
massive stars in each region. Their spatial distribution is a fossil imprint of
recently completed star formation, providing much needed constraints for models
of molecular cloud and cluster formation and dissipation;
with velocity dispersions of $\sim$1\,km\,s$^{-1}$ (e.g. {\it de Bruijne}, 1999)
the stars simply have not traveled far from their birth sites ($\sim 10$ pc in
10 Myr).
Low-mass stars also provide better kinematics,
because it is easier to obtain accurate radial velocities from the many metallic lines
in G, K and M type stars than it is from O and B type stars. 

\bigskip
\noindent
\section{\textbf{SEARCHES FOR LOW-MASS PMS
STARS IN OB ASSOCIATIONS}
}
\bigskip

Except for the youngest, mostly embedded populations in the molecular clouds,
or dense, optically visible clusters like the Orion Nebula Cluster (ONC),
most of the low-mass stellar population in nearby OB associations is widely
spread over tens or even hundreds of square degrees on the sky. 
Moreover, it is likely
that after $\sim 4$ Myr the stars are no longer associated with their parent
molecular clouds,
making it difficult to sort them out from the field population.
Therefore, a particular combination of various instruments and techniques
is required to reliably single out the low-mass PMS stars.
The main strategies that have been used to identify these populations are
objective prism surveys, X-ray emission, proper motions and, more recently,
variability surveys.

\bigskip
\noindent
\textbf{ 2.1 Objective Prism Surveys}
\bigskip

The TTS originally were identified as stars of late spectral types (G-M),
with strong emission lines (especially H$\alpha$) and erratic light
variations, spatially associated with regions of dark nebulosity ({\it Joy}, 1945).
Stars resembling the original variables first identified as TTS
are currently called "strong emission" or Classical TTS (CTTS).
Subsequent spectroscopic studies of the Ca II H and K lines and the first X-ray
observations with the {\it Einstein} X-ray observatory 
({\it Feigelson and De Campli}, 1981; {\it Walter and Kuhi}, 1981)
revealed surprisingly strong X-ray activity in TTS, 
exceeding the solar levels by several orders of
magnitude, and also revealed a population of X-ray strong objects
lacking the optical signposts of CTTS, like strong
H$\alpha$ emission.  These stars, initially called "naked-T Tauri
stars" ({\it Walter and Myers}, 1986), are now widely known as "weak-line" TTS after 
{\it Herbig and Bell} (1988). The CTTS/WTTS dividing line was set at W(H$\alpha$) $= 10$\AA. 
In general, the excess H$\alpha$
emission in WTTS seems to originate in enhanced solar-type magnetic activity
({\it Walter et al.}, 1988), while the extreme levels observed in CTTS can be explained by
a combination of enhanced chromospheric activity and emission coming from
accretion shocks in which material from a circumstellar disk is funneled along
magnetic field lines onto the stellar photosphere (Section~5).
Recently, {\it White and Basri} (2003) revisited the WTTS/CTTS classification
and suggested a modified criterion that takes into account the contrast
effect in H$\alpha$ emission as a function of spectral type in stars cooler
than late K. \\

The strong H$\alpha$ emission characteristic of low-mass young stars, and 
in particular of CTTS, encouraged early 
large scale searches using photographic plates and objective prisms
on wide field instruments like Schmidt telescopes
(e.g., {\it Sanduleak}, 1971, in Orion).
These very low resolution spectroscopic surveys (typical dispersions
of $\sim 500$\AA/mm to $\sim 1700$ \AA/mm at H$\alpha$;
c.f. {\it Wilking et al.}, 1987; {\it Brice\~no et al.}, 1993)
provided large area coverage, allowed estimates of
spectral types and a qualitative assessment of the strength of 
prominent emission lines, like the hydrogen Balmer lines or the Ca II H \& K lines.
{\it Liu et al.} (1981) explored a $5\times 5$ deg region in Per OB2 and detected
25 candidate TTS. {\it Ogura} (1984) used the 1m Kiso Schmidt to find 135 H$\alpha$
emitting stars in Mon OB1. {\it Wilking et al.} (1987) detected 86 emission line objects
over 40 square degrees in the $\rho$ Ophiuchi complex.
{\it Mikami and Ogura} (2001) searched an area of 36 square degrees in Cep OB3
and identified 68 new emission line sources.
In the Orion OB1 association, the most systematic search was that done with
the 1m Kiso Schmidt (e.g., {\it Wiramihardja et al.}, 1989, 1993),
covering roughly 150 square degrees and
detecting $\sim 1200$ emission line stars, many of which were argued to be
likely TTS.
{\it Weaver and Babcock} (2004) recently identified 63 H$\alpha$ emitting objects
in a deep objective prism survey of the $\sigma$ Orionis region.

The main limitation of this technique is the strong bias towards
H$\alpha$-strong PMS stars; few WTTS can be detected at the resolution
of objective prisms (c.f. {\it Brice\~no et al.}, 1999). 
{\it Brice\~no et al.} (2001) find that only
38\% of the 151 Kiso H$\alpha$ sources falling within their 
$\sim 34$ square degree survey area in the Orion OB 1a and 1b sub-associations
are located above the ZAMS in color-magnitude diagrams, and argue
that the Kiso survey is strongly contaminated by foreground main sequence
stars (largely dMe stars). The spatial distribution of the Kiso sources 
has been useful to outline the youngest regions in Orion, where the highest 
concentrations of CTTS are located ({\it G\'omez and Lada}, 1998), 
but these samples can be dominated by field stars in regions far from 
the molecular clouds, in which  the CTTS/WTTS fraction is small.
Therefore, as with other survey techniques,
objective prism studies require follow up spectroscopy to confirm membership.

\bigskip
\noindent
\textbf{ 2.2 X-ray surveys}
\bigskip

Young stars in all evolutionary stages, from
class~I protostars to ZAMS stars, show strong X-ray
activity (for recent reviews on the X-ray properties of YSOs
see {\it Feigelson and Montmerle}, 1999 and
{\it Favata and Micela}, 2003).
After the initial Einstein studies, the ROSAT and ASCA X-ray observatories
increased considerably the number of observed star forming regions, and thereby the
number of known X-ray emitting TTS.
Today, XMM-Newton and $Chandra$ 
allow X-ray studies of star forming regions at unprecedented
sensitivity and spatial resolution.

X-ray observations are a well established tool
to find young stars. For nearby OB associations, which typically cover
areas in the sky much larger than the field-of-view of X-ray observatories,
deep and spatially complete observations are usually not feasible.
However, large scale shallow surveys have been conducted
with great success.
The ROSAT All Sky Survey (RASS) provided coverage of the
whole sky
in the $0.1-2.4$~keV soft X-ray band. With a mean limiting
flux of about $2 \times 10^{-13}\rm\,erg\> s^{-1}\,cm^2$
this survey provided
a spatially complete, flux-limited sample of X-ray sources
that led to the
detection of hundreds of candidate PMS stars in star forming
regions all over the sky (see {\it Neuh\"auser}, 1997).

The X-ray luminosities of young stars for a given age, mass, and
bolometric luminosity can differ by several orders of magnitude.
Until recently it was not even clear whether all young stars are
highly X-ray active, or whether an ``X-ray quiet''
population of stars with suppressed magnetic activity may exist,
which would have introduced a serious bias in any X-ray selected
sample.
The $Chandra$ Orion Ultradeep Project ({\it Getman et al.}, 2005),
a 10 day long
observation of the Orion Nebular Cluster, has provided
the most comprehensive dataset ever acquired on the X-ray
emission of PMS, and solved this question
by providing definitive information
on the distribution of X-ray luminosities in young stars.
It found no indications for ``X-ray quiet'' TTS, and established that
50\% of the TTS have $\log\left(L_{\rm X}/L_{\rm bol}\right) \ge -3.5$,
while 90\% have $\log\left(L_{\rm X}/L_{\rm bol}\right) \ge -4.5$
({\it Preibisch et al.}, 2005; also see chapter by {\it Feigelson et al.}).
Since the RASS flux limit corresponds to X-ray luminosities
of about $\rm 5 \times 10^{29}\>erg\> s^{-1}$ at the distance of the nearest
OB associations ($\sim 140$ pc),
this implies that the RASS data
are essentially complete
only for $M \ge 1~\msun$ PMS stars in those regions, while only a
fraction of the X-ray brightest sub-solar mass PMS stars are detected.
A caveat of the RASS surveys for PMS stars is that these samples
can be significantly contaminated by foreground, X-ray
active zero age main sequence stars ({\it Brice\~no et al.}, 1997).
These limitations have to be kept in mind when working whith X-ray
selected samples; at any rate, follow-up observations are necessary
to determine the nature of the objects.

\bigskip
\noindent
\textbf{ 2.3 Proper Motion surveys.}
\bigskip

The recent availability of ever-deeper, all-sky catalogs
of proper motions (like {\it Hipparcos} and the Tycho family of
catalogs)
has aided the effort in identifying the low-mass
members of the nearest OB associations. The proper
motions of members of a few of the nearest OB associations
are of the order of tens of mas\,yr$^{-1}$
({\it de Zeeuw et al.}, 1999). With proper motions whose errors
are less than a few\,mas\,yr$^{-1}$, one can attempt to
kinematically select low-mass members of nearby associations.
The nearest OB association, Sco-Cen, has been the most
fruitful hunting ground for identifying low-mass members
by virtue of their proper motions. Current proper motion
catalogs (e.g., Tycho-2, UCAC) are probably adequate to
consider kinematic selection of low-mass stars in at least
a few other nearby groups (e.g., Vel OB2, Tr 10, $\alpha$ Per,
Cas-Tau, Cep OB6). The very small ($<$10\,mas\,yr$^{-1}$)
proper motions for some of the other nearby OB associations
(e.g., Ori OB1, Lac OB1, Col 121) will preclude
any attempts at efficient selection of low-mass members
via proper motions, at least with contemporary astrometric
catalogs.

The {\it Hipparcos}
survey of the nearest OB associations by {\it de Zeeuw et al.} (1999)
was able to identify dozens of FGK-type stars as candidate
members. {\it De Zeeuw et al.} (1999) predicted
that $\sim$37/52 (71\%) of their GK-type {\it Hipparcos} candidates
would be bona fide association members, and indeed
{\it Mamajek et al.} (2002) found that 22/30 (73\%) of a subsample of
candidates, located in Sco-Cen,
 could be spectroscopically confirmed as PMS
stars. {\it Hoogerwerf} (2000) used the ACT and TRC
proper motion catalogs (p.m. errors $\simeq$ 3\,mas\,yr$^{-1}$)
to identify thousands of candidate
Sco-Cen members down to V $\sim 12$.
Unfortunately, the vast majority of stars in the ACT and TRC
catalogs, and their descendant (Tycho-2), do not have
known spectral types or parallaxes (in contrast to the
{\it Hipparcos} catalog), and hence the contamination level is
large. {\it Mamajek et al.} (2002) conducted a spectroscopic
survey of an X-ray and color-magnitude-selected
subsample of the Hoogerwerf proper motion-selected sample
and found that 93\% of the candidates were bona fide
PMS association members.
The high quality proper motions also enabled the estimate
of individual parallaxes to the Upper Centaurus Lupus (UCL) and 
Lower Centarus Crux (LCC) members, reducing
the scatter in the HR diagram ({\it Mamajek et al.}, 2002).
In a survey of 115 candidate Upper Sco members selected
solely via STARNET proper motions (p.m. errors of
$\sim$5\,mas\,yr$^{-1}$), {\it Preibisch et al.} (1998) found
that {\it none} were PMS stars. The lesson learned
appears to be that proper motions {\it alone} are insufficient for
efficiently identifying low-mass members of nearby OB associations.
However, when proper motions are used in conjunction with
color-magnitude, X-ray, spectral type, or parallax data
(or some combination thereof), finding low-mass
associations can be a very efficient task.

\bigskip
\noindent
\textbf{ 2.4 Photometric surveys: Single-epoch observations}
\bigskip

Single epoch photometric surveys are
frequently used to select candidate low-mass members of young clusters
or associations.  
Most studies use broadband, optical filters that are
sensitive to the temperatures of G, K and M-type stars.  Near-IR color-magnitude
diagrams (CMDs)
are not as useful for selecting low-mass PMS stars because NIR colors
are similar for all late type stars. 

\begin{figure}[ht]
\epsscale{1.0}
\plotone{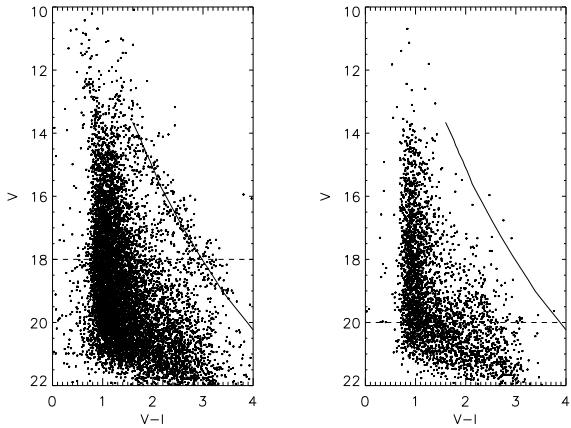}
\caption{\small The left panel shows the V vs. V$-$I$_C$ color-magnitude
  diagram of 9556 stars in 0.89 deg$^2$ around $\sigma$~Ori
  (from {\it Sherry et al.}, 2004).  The
  solid line is a 2.5~Myr isochrone ({\it Baraffe et al.}, 1998; {\it Baraffe et
  al.}, 2001) at a distance of 440~pc ({\it de Zeeuw et al.}, 1999), extending from
  $1.2~\msun$ at V$\sim 13.5$ 
  down to $\sim 0.2~\msun$ at V$\sim 18$ (the completeness limit, indicated
  by the dashed line).  
  This isochrone marks the expected position of the PMS locus for Orion
  OB1b.  There is a clear increase in the density of stars around the
  expected position of the PMS locus.  
  The right panel shows the same
  color-magnitude diagram (CMD) for the 0.27~deg$^2$ control fields from {\it Sherry et al.} (2004).
  The isochrone (solid line)
  is the same as in the left panel.  The dashed line marks the fainter
  completeness limit of the control fields.
  }
\label{sherry_cmd}
\end{figure}

Candidate low-mass association members are usually selected by their
location in the CMD above the zero-age main sequence (ZAMS).
This locus is usually defined by either a known (spectroscopically
confirmed) population of PMS stars
or because the PMS population of the association
is clearly visible as a concentration on the CMD (e.g.,
Fig.~1).
Single epoch photometry is most effective in regions such as
$\sigma$~Ori or the ONC where the proximity and youth of the cluster
make the low-mass PMS members brighter than the bulk of the field
stars at the colors of K and M-type stars.

The main advantage of photometric selection is that for a specified amount
of time on any given telescope a region of the sky can be surveyed to
a fainter limit than can be done by a variability survey or a
spectroscopic survey.  Also, photometric selection can 
identify low-mass association members with very low amplitude
variability. 
The disadvantage of single epoch photometric selection is that there
is inevitably some contamination by
foreground field stars and background giants.  As with the other techniques, 
it is impossible to securely identify any individual star as a low-mass
member of the association without spectroscopic follow-up.  In small areas
with a high density of low-mass association members such as
the $\sigma$~Ori cluster or the ONC, single epoch photometry can
effectively select the low-mass population because field star
contamination is fairly small ({\it Sherry et al.}, 2004; {\it Kenyon et al.},
2005).  But in large areas with a lower density of low-mass association members,
such as Orion OB1b (Orion's belt) or Orion OB1a (NW of the belt) the
field star contamination can be large enough to make it difficult to
even see the PMS locus.

\bigskip
\noindent
\textbf{ 2.5 Photometric surveys: Variability}
\bigskip

 Variability in  T Tauri stars has  been intensively
studied  over the years (e.g., {\it Herbst et al.}, 1994), but mostly 
as follow up observations of individual 
young stars that had been identified by some other means. 
Building on the availability of large format CCD cameras installed on
wide-field telescopes it has now become feasible to conduct multi-epoch,
photometric surveys that use variability to pick out candidate 
TTS over the extended areas spanned by nearby OB associations.
In Orion, two major studies have been conducted over the past few years.
{\it Brice\~no et al.} (2001, 2005a) have done a VRI variability survey
using the Quest I CCD Mosaic Camera installed on the Venezuela 1m Schmidt,
over an area of $\simgreat 150$ square degrees in the Orion OB1 
association. In their first release, based on some 25 epochs  
and spanning an area of 34 square degrees, they 
identified  $\sim 200$ new, spectroscopically confirmed, 
low-mass members, and a new 10 Myr old 
clustering of stars around the star 25 Ori ({\it Brice\~no et al.}, 2006). 
{\it McGehee et al.} (2005) analysed 9 repeated observations over 25 square
degrees in Orion, obtained with the Sloan Digital Sky Survey (SDSS). 
They selected 507 stars that met their variability criterion in the SDSS g-band. 
They did not obtain follow  up spectra of their candidates, rather, 
they apply their observations in a statistical sense to search for photometric 
accretion-related signatures in their lower mass candidate members.  
{\it Slesnick et al.} (2005) are using the  QUEST II CCD Mosaic Camera on the
Palomar Schmidt to conduct a BRI, multi-epoch survey of $\sim 200$ square
degrees in Upper Sco, 
and {\it Carpenter et al.} (2001) has used  repeated observations made  with 2MASS  in a
$0.86 \times 6$ degree strip centered on  the  ONC  to study the near-IR
variability of a large sample of young, low-mass stars.  

With more wide angle detectors on small and medium sized telescopes,
and projects like LSST coming on line within less than a decade, variability 
promises to grow as an efficient means of selecting large samples of
candidate low-mass PMS stars down to much lower masses than available to
all-sky X-ray surveys like ROSAT, and without the bias towards CTTS of 
objective prism studies. However, as with every technique there are 
limitations involved, e.g., temporal sampling and a bias toward
variables with larger amplitudes, especially at the faint end, are 
issues that need to be explored.

\bigskip
\noindent
\textbf{ 2.6 Spectroscopy.}
\bigskip

As already emphasized in the previous paragraphs, low to moderate resolution 
follow-up spectroscopy is essential to confirm membership of PMS stars. 
However, the observational effort to identify
the widespread population of PMS stars among the many thousands of
field stars in the large areas spanned by nearby OB associations
is huge, and up to recently has largely precluded further investigations.
With the advent of extremely powerful multiple-object spectrographs, such
as 2dF at the Anglo-Australian Telescope,
Hydra on the WIYN 3.5~m and the CTIO 4~m telescopes,
and now Hectospec on the 6.5~m MMT, 
large scale spectroscopic surveys have now become feasible.

One of the most powerful approaches to unbiased surveys of 
young low-mass stars is to use the 
presence of strong \mbox{6708\,\AA}\,Li I absorption lines as a 
diagnostic of the PMS nature of a candidate object 
(e.g., {\it Dolan and Mathieu}, 1999, 2001).
Because Li I is strongly diminished in very early phases of stellar
evolution, a high Li content is a reliable indication for the youth
of a star (e.g., {\it Herbig}, 1962; {\it D'Antona and Mazzitelli}, 1994).
However, Li depletion is not only a function of stellar age, but
also of stellar mass and presumably even depends on additional
factors like stellar rotation (cf. {\it Soderblom}, 1996).
Not only PMS stars, but also somewhat older, though still relatively young
zero-age main sequence stars, e.g.~the
G and K type stars in the $\sim 10^8$ years old Pleiades
(c.f. {\it Soderblom et al.} 1993), can display  Li absorption lines.
In order to classify stars as PMS,
one thus has to consider a spectral type dependent threshold for the
Li line width.
Such a threshold can be defined by the upper envelope of Li measurements
in young clusters of main-sequence stars with ages between $\sim\!30$ Myr
and a few $100$ Myrs such as
IC 2602, IC 4665, IC 2391, $\alpha$ Per, and the Pleiades.
Any star with a Li line width considerably above this threshold
should be younger than $\sim\!30$ Myrs and can therefore be
classified as a PMS object.

In addition to Li, other spectroscopic signatures can be used as
youth indicators, such as the K I and Na I absorption lines 
that are typically weaker in low-mass PMS objects compared to field
M-type dwarfs ({\it Mart{\'\i}n et al.}, 1996; {\it Luhman}, 1999),
and radial velocities (if high resolution spectra are available).

\bigskip
\noindent
\section{\textbf{AGES AND AGE SPREADS OF LOW-MASS STARS}}
\bigskip

Stellar ages are usually inferred from the positions of the stars
in the HR-diagram by comparison with theoretical PMS evolutionary models.
Photometry of young clusters and associations has shown that low-mass
members occupy a wide swath on the CMD or the H-R~diagram (NGC 2264;
ONC; $\sigma$~Ori).  A common interpretation has been that this spread
is evidence of an age spread among cluster members (e.g.,
{\it Palla and Stahler}, 2000). This interpretation assumes
that the single epoch observed colors and magnitudes of low-mass PMS
stars accurately correspond to the temperatures and luminosities of
each star.  
However, it has to be strongly emphasized that the masses and especially
the ages of the individual stars read off from their position in these diagrams
are generally {\em not} identical to their true masses and ages.
Several factors can cause considerable deviations of an individual star's
position in the HR-diagram from the locations predicted by theoretical
models for a given age and mass.
Low-mass PMS stars exhibit variability
ranging from a few tenths in the WTTS up to several magnitudes in 
the CTTS ({\it Herbst et al.}, 1994).  Furthermore, binarity can add a
factor of 2 ($\sim 0.75$ mag) spread to the CMD. 
For regions spanning large areas on the sky
there will be an additional spread
caused by the distribution of distances along the line of sight
among individual stars.
Another source of uncertainty are the
calibrations used to derive bolometric luminosities 
and effective temperatures, and finally, the choice of
evolutionary tracks used, that in some cases yield 
different ages for intermediate ($\sim 2-5~\msun$) 
and low-mass stars located in the same
region (e.g., the ONC; {\it Hillenbrand}, 1997; see also
{\it Hartmann}, 2001).
Therefore,
any age estimates in young star forming regions
must account for the significant spread that a coeval population
will necessarily have on the CMD, which translates into a spread on the
H-R~diagram.

\begin{figure}[ht]
\epsscale{1.0}
\plotone{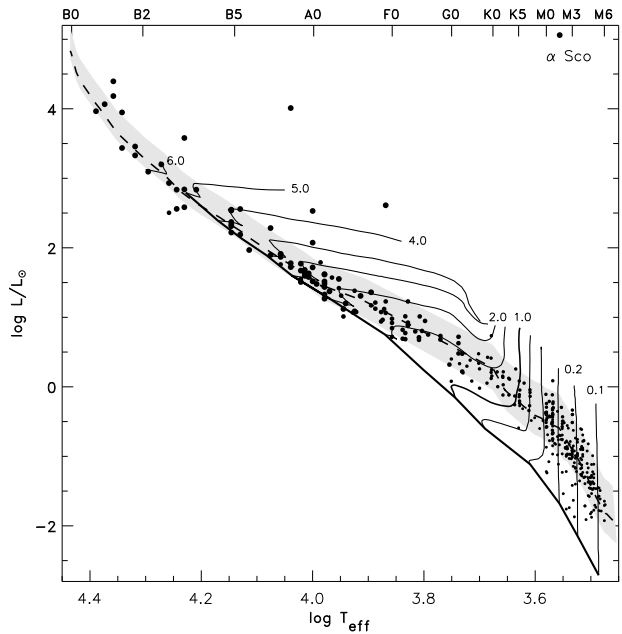}
\caption{\small
HR diagram for the Upper Sco association members 
from the study of {\it Preibisch et al.} (2002).  The lines show the
evolutionary tracks from the {\it Palla and Stahler} (1999) PMS models, some labeled
by their masses in solar units.  The thick solid line shows the main sequence.
The 5 Myr isochrone is shown as the dashed line; it was composed from the
high-mass isochrone from {\it Bertelli et al.} (1994) for masses
$6\!-\!30\,M_\odot$, the {\it Palla and Stahler} (1999) PMS models for masses
$1\!-\!6\,M_\odot$, and the {\it Baraffe et al.} (1998) PMS models for masses
$0.02\!-\!1\,M_\odot$.
The grey shaded band shows the region in which one expects 90\%
of the member stars to lie, based on the assumption of a common age of
5 Myr for all stars and taking proper account of the uncertainties
and the effects of unresolved binaries (for details see text).
}
\label{preibisch_hrd}
\end{figure}

As an example,
Fig.~2 shows the HR diagram containing all Upper Sco
association members from {\it Preibisch et al.} (2002);
the diagram also shows the main sequence and a 5~Myr isochrone. 
Not only the majority of the low-mass stars, but also most of the
intermediate- and high-mass stars lie close to or on the 5~Myr
isochrone. There clearly is
a considerable scatter that may seem to suggest
a spread in stellar ages. 
In the particular case of Upper Sco shown here,
in addition to the other effects mentioned above,
the most important factor for the apparent scatter
is the relatively large
spread of individual stellar distances ($\sim \pm\,20$~pc around the
mean value of 145 pc; {\it de Bruijne}, 1999 and
priv.~comm.) in this very nearby and extended region, which
causes the luminosities to be either over or under estimated
when a single distance is adopted for all sources.
A detailed discussion and statistical modeling of these effects
is given in {\it Preibisch and Zinnecker} (1999) and {\it Preibisch et al.} (2002).  
In the later work these authors showed that
the observed HR diagram for the low-mass stars in Upper Sco
is consistent with the assumption of
a {\em common stellar age of about 5 Myr; there is no evidence
for an age dispersion, although small
ages spreads of $\sim 1-2$~Myr cannot be excluded by the data}.
{\it Preibisch et al.} (2002) showed that the derived age is also robust
when taking into account
the uncertainties of the theoretical PMS models.
It is remarkable that the isochronal age derived for the low-mass stars
is consistent with
previous and independent age determinations based on the nuclear and
kinematic ages of the massive stars ({\it de Zeeuw and Brand}, 1985; {\it de Geus et al.}, 1989),
which also yielded 5 Myr.
This very good agreement of the {\em independent\/} age determinations for the
high-mass and the low-mass stellar population shows that {\em low- and
high-mass stars are coeval} and thus have formed together.  Furthermore, the
absence of a significant age dispersion implies that all stars in the
association have formed more or less simultaneously. Therefore, the
star-formation process must have started rather suddenly and everywhere at the
same time in the association, and also must have ended after
at most a few Myr.  The star formation process in Upper Sco can thus be
considered as a {\em burst of star formation\/}.

{\it Sherry} (2003) compared the observed spread across the V vs. V$-$I$_C$
CMD of low-mass members of the $\sigma$~Ori cluster to the spread that
would be expected based upon the known variability of WTTS, the field
binary fraction, the photometric errors of his survey, and a range of
simple star formation histories.  The observed spread was consistent
with the predicted spread of an isochronal population.  Sherry 
concluded that the bulk of the low-mass stars must have formed over a
period of less than $\sim 1$ Myr.  A population with a larger age spread
would have been distributed over a larger region of the CMD than was
observed.  

{\it Burningham et al.} (2005) also examined the possibility of an age spread
among members of the $\sigma$~Ori cluster.  They used two epoch R and
i' observations of cluster members taken in 1999 and 2003 to estimate
the variability of each cluster member.  They then constructed a series of
simple models with a varying fraction of equal mass binaries.  They
found that the observed spread on the CMD was too large to be fully
accounted for by the combined effects of observational errors,
variability (over 1-4 years), and binaries.  They conclude that the
larger spread on the CMD could be accounted for by either a longer period
of accretion driven variability, an age spread of $\la 2$ Myrs (using a
distance of 440~pc or 4~Myrs using a distance of 350~pc), or a
combination of long term variability and a smaller age spread.  
However, their result is not necessarily in contradiction with the findings of
{\it Sherry} (2003), especially if the actual variability is larger than
they estimate based on their two epoch observations.

\begin{figure}[ht]
 \epsscale{1.0}
\plotone{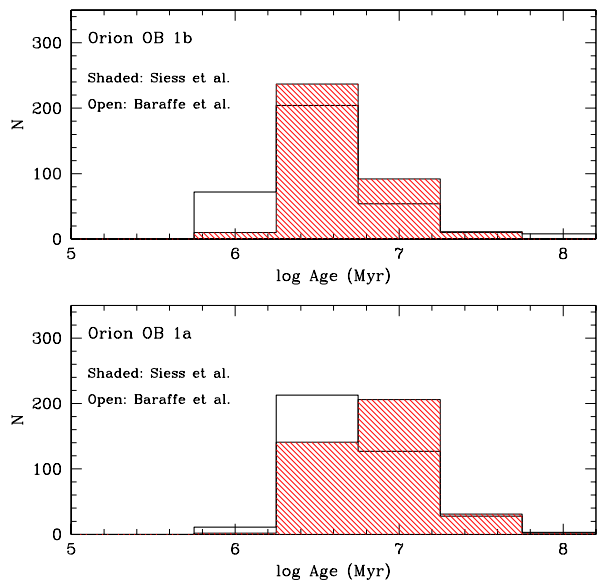}
 \caption{\small Distribution of ages of $\sim 1000$
newly identified TTS in a $\sim 60$ square degree area
spanning the the Orion OB 1a and 1b
sub-associations ({\it Brice\~no et al.}, in preparation).
Ages were derived using the
{\it Baraffe et al.} (1998) and {\it Siess et al.} (2000)
evolutionary tracks and isochrones. The mean ages are
$\sim 4$ Myr for Ori OB 1b and $\sim 8$ Myr for Ori OB 1a.
}
\label{ori_age_histo}
\end{figure}

In the Orion OB1 association, the V and $\rm I_C$-band CMDs 
for spectroscopically confirmed TTS ({\it Brice\~no et al.}, 2005a, 2006), 
which mitigate the spread caused by the
variability of individual sources by plotting their mean magnitudes
and colors, provide evidence that the Ori OB 1b sub-association is 
older ($\sim 4$ Myr) than the often quoted age of about 1.7 Myr 
({\it Brown et al.}, 1994),
with a rather narrow age distribution regardless of the tracks used
(Fig.~3). The same dataset for the older OB 1a sub-association also shows a
narrow range of ages, with a mean value of $\sim 8$ Myr.

The $\lambda$ Ori region also shows that the age distribution and star formation  
history is spatially dependent. {\it Dolan and Mathieu} (2001; see also {\it Lee et al.}, 2005) 
found that age distributions of high-mass and solar-type stars in the region show 
several critical features:

a) Both high- and low-mass star formation began concurrently in the center of the 
SFR roughly 6-8 Myr ago;

b) Low-mass star formation ended in the vicinity of $\lambda$ Ori roughly 1 Myr ago;

c) Low-mass star formation rates near the B30 and B35 clouds reached their maxima later 
than did low-mass star formation in the vicinity of $\lambda$ Ori;

d) Low-mass star formation continues today near the B30 and B35 clouds.

As with the Ori OB1 associations, this varied star-formation history reflects 
the rich interplay of the massive stars and the gas.

The accumulated evidence for little, if any, age spreads in various
star forming regions provides a natural explanation for the
"post-T Tauri problem" (the absence of "older" TTS in star forming
regions like Taurus, assumed to have been forming stars for up to 
tens of Myrs, {\it Herbig}, [1978]), and at the same time implies 
relatively short lifetimes for molecular clouds. 
These observational arguments support the
evolving picture of star formation as a fast and remarkably
synchronized process in molecular clouds (Section~6).

\bigskip
\noindent
\section{\textbf{THE IMF IN OB ASSOCIATIONS}}
\bigskip

The IMF is the utmost challenge for any theory
of star formation.
Some theories suggest that the IMF should
vary systematically with the star formation environment ({\it Larson}, 1985), 
and for many years star formation was supposed to be a bimodal
process (e.g., {\it Shu and Lizano}, 1988) according to which high- and
low-mass stars should form in totally different sites.
For example, is was suggested that increased heating due to the strong
radiation from massive stars raises the Jeans mass,
so that the bottom of the IMF
would be truncated in regions of high-mass star formation.
In contrast to the rather quiescent environment in small low-mass clusters
and T associations (like the Taurus molecular clouds),
forming stars in OB association are exposed to the
strong winds and intense UV radiation of the massive stars,
and, after a few Myr, also affected by supernova explosions.
In such an environment, it may be harder to form low-mass stars, because,
e.g., the lower-mass cloud cores may be
completely dispersed before protostars can even begin to form.

Although it has been long established that low-mass stars can form alongside
their high-mass siblings (e.g., {\it Herbig}, 1962) in nearby OB associations,
until recently it was not well known
what quantities of low-mass stars are produced in OB environments.
If the IMF in OB associations is not truncated and
similar to the
field IMF, it would follow that most of their total stellar mass $(\ga 60\%)$
is found in low-mass $(< 2\,M_\odot)$ stars.
This would then imply that most of the current galactic star formation
is taking place in OB associations.
Therefore, the typical environment for forming stars (and planets)
would be close to massive stars and not in isolated regions like Taurus.

In Fig.~4 we show the empirical mass function for Upper Sco
as derived in {\it Preibisch et al.} (2002), 
for a total sample of 364 stars covering the mass range from
$0.1\,M_\odot$ up to $20\,M_\odot$.
The best-fit multi-part power law function for the
probability density distribution is given by
\begin{equation}
\frac{{\rm d}N}{{\rm d}M} \propto   \left\{ \begin{array}{lcrcr}
     M^{-0.9\pm0.2} & \mbox{for} &0.1\!&\le M/M_\odot <&\!0.6\\
     M^{-2.8\pm0.5} & \mbox{for} &0.6\!&\le M/M_\odot <&\!2\\
     M^{-2.6\pm0.5} & \mbox{for} &2\!&\le M/M_\odot <&\!20\end{array}\right.
 \label{imf_usco}
\end{equation}
or, in shorter notation,
$\alpha[0.1\!-\!0.6] = -0.9\pm0.2$,
$\alpha[0.6\!-\!2.0] = -2.8\pm0.5$,
$\alpha[2.0\!-\!20] = -2.6\pm0.3$.
For comparison, the plot also shows two different field IMF models,
the {\it Scalo} (1998) model, which is given by
$\alpha[0.1\!-\!1] = -1.2\pm0.3$, $\alpha[1\!-\!10] = -2.7\pm0.5$,
$\alpha[10\!-\!100] = -2.3\pm0.5$, and
the {\it Kroupa} (2002) model with
$\alpha[0.02\!-\!0.08] = -0.3\pm0.7$, $\alpha[0.08\!-\!0.5] =
-1.3\pm0.5$, $\alpha[0.5\!-\!100] = -2.3\pm0.3$.

\begin{figure}[ht]
\epsscale{1.0}
\plotone{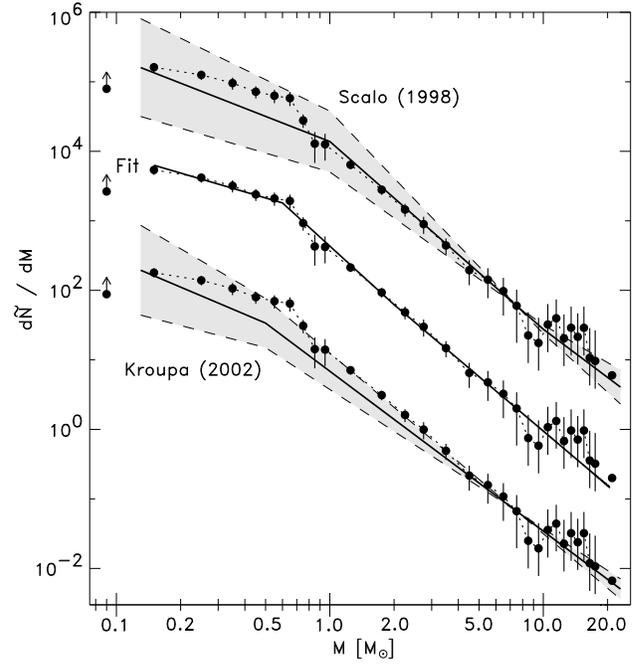}
\caption{\small
Comparison of the mass function derived for the Upper Sco association
with different mass function measurements for the field 
(from {\it Preibisch et al.}, 2002).
The Upper Sco mass function is shown three times by the solid dots
connected with the dotted lines, multiplied by arbitrary factors. The middle
curve shows the original mass function, the solid line is our multi-part
power-law fit.  The upper curve shows our mass function multiplied by a factor
of 30 and compared to the {\it Scalo (1998)} IMF (solid line); the grey shaded area
delimited by the dashed lines represents the range allowed by the errors of
the model.  The lower curve shows our mass function multiplied by a factor of
1/30 and compared to the {\it Kroupa} (2002) IMF (solid line).
}
\label{preibisch_imf}
\end{figure}

While the slopes of the fit to the empirical mass function of Upper Sco
are not identical to
those of these models, they are well
within the ranges of slopes derived for similar mass ranges in other young
clusters or associations, as compiled in {\it Kroupa} (2002).  Therefore,
it can be concluded that,
within the uncertainties, the general shape of the Upper Sco
 mass function is consistent with recent field star and cluster IMF
determinations.

In the Orion OB1 association {\it Sherry et al.} (2004) find that the mass function
for the $\sigma$ Ori cluster is consistent with the {\it Kroupa} (2002) IMF.
In the immediate vicinity of $\lambda$ Ori, {\it Barrado y Navascues et al.} (2004) combined 
their deep imaging data with the surveys of {\it Dolan and Mathieu} (1999, 2001; 
limited to the same area) to obtain an initial mass function from $0.02 - 1.2~\msun$. 
They find that the data indicate a power law index of $\alpha = -0.60\pm 0.06$ across the 
stellar-substellar limit and a slightly steeper index of $\alpha = -0.86 \pm 0.05$ over the 
larger mass range of $0.024~\msun$ to $0.86~\msun$, much as is found in other young regions.

Over the entire $\lambda$ Ori star-forming region, {\it Dolan and Mathieu} (2001) were able to 
clearly show that the IMF has a spatial dependence. {\it Dolan and Mathieu} (1999) had found 
that within the central $\sim 3.5$~pc around $\lambda$ Ori the low-mass 
stars were deficient by a factor of 2 compared to the field IMF. Outside this central field, 
{\it Dolan and Mathieu} (2001) showed the low-mass stars to be overrepresented compared to the 
{\it Miller and Scalo} (1978) IMF by a factor of 3. A similar over-representation of low-mass stars 
is also found at significant confidence levels when considering only stars associated 
with B30 and B35.

Thus the global IMF of the $\lambda$ Ori SFR resembles the field, while the local IMF 
appears to vary substantially across the region. No one place in the $\lambda$ Ori SFR 
creates the field IMF by itself. Only the integration of the star-formation process 
over the entire region produces the field IMF.

\bigskip
\noindent
\section{\textbf{DISK EVOLUTION}}
\bigskip

The presence of circumstellar disks around low-mass
pre-main sequence stars appears to be a natural 
consequence of the star formation process;
these disks play an important role both in determining the final 
mass of the star and as the potential sites for planet formation.
Though we have a good general understanding of the overall 
processes involved, many important gaps still remain.
For instance, the times scales for mass accretion and disk 
dissipation are still matters of debate. An example is the 
discovery of a seemingly long lived accreting disk around 
the $\sim 25$ Myr old, late type (M3) star {\em St 34} located 
in the general area of the Taurus dark clouds 
({\it Hartmann et al.}, 2005a; {\it White and Hillenbrand}, 2005).

How circumstellar disks evolve and  whether their
evolution is affected by environmental conditions
are questions that at present can only be investigated by 
looking to low-mass PMS stars, and in particular the best
samples can now be drawn from the various nearby OB associations. 
First, low-mass young stars like T Tauri stars, in particular 
those with masses $\sim 1 \> \msun$ constitute good 
analogues of what the conditions may have been in the
early Solar System. 
Second, as we have discussed in Section~3, OB associations 
can harbor many stellar aggregates, with distinct ages,
such as in Orion OB1, possibly the result of 
of star-forming events (triggered or not) occurring 
at various times throughout the original GMC. 
Some events will have produced dense clusters while
others are responsible for the more spread out population. 
The most recent events are easily recognizable by the
very young ($\simless 1$ Myr) stars still embedded in their 
natal gas, while older ones may be traced by the $\sim 10$  Myr
stars which have long dissipated their parent clouds. 
This is why these regions provide 
large numbers of PMS stars in different environments, 
but presumably sharing the same "genetic pool",
that can allow us to build a {\sl differential picture} of how disks
evolve from one stage to the next.

Disks are related to many of the photometric and spectroscopic 
features observed in T Tauri stars.
The IR emission originates by the contribution
from warm dust in the disk, heated at a range of temperatures
by irradiation from the star and viscous dissipation (e.g., {\it Meyer et al.}, 1997).
The UV excesses, excess continuum emission (veiling), irregular photometric variability, 
broadened spectral line profiles (particularly in the hydrogen lines and others like Ca II),
are explained as different manifestations of gas accretion from a circumstellar
disk. In the standard magnetospheric model ({\it K\"onigl}, 1991), the accretion disk is truncated
at a few stellar radii by the magnetic field of the star, the disk material
falls onto the photosphere along magnetic field lines at supersonic velocities, creating
an accretion shock which is thought to be
largely responsible for the excess UV and continuum emission ({\it Calvet and Gullbring}, 1998).
The infalling material also produces the observed broadened and P Cygni profiles
observed in hydrogen lines ({\it Muzerolle et al.}, 1998a,b, 2001). 
Disk accretion rates for most CTTS are of the order of $\rm 10^{-8} \> \msun \> yr^{-1}$ at
ages of a 1-2 Myr (e.g., {\it Gullbring et al.}, 1998; {\it Hartmann}, 1998; {\it Johns-Krull and Valenti}, 2001).

Comparative studies of near-IR emission 
and accretion-related indicators (H$\alpha$ and Ca II emission, UV excess
emission) at ages $\sim 1-10$ Myr offer
insight into how the innermost part of the disk evolves. 
One way to derive the fraction of stellar systems with inner disks
is counting the number of objects showing excess emission in the JHKL near-IR bands. 
The availability of 2MASS has made JHK studies of young populations over wide spatial
scales feasible.
More recently, the emission of T Tauri stars in the Spitzer IRAC and MIPS bands
has been characterized by {\it Allen et al.} (2004) and 
{\it Hartmann  et al.} (2005b).
Another approach is determining the number of objects that
exhibit strong H$\alpha$ and Ca II emission (CTTS), or
UV excesses; these figures provide an indication of how many systems
are actively accreting from their disks.

So far, the most extensive studies of how disk fractions change with
time have been conducted in the Orion OB1 association.
{\it Hillenbrand et al.} (1998) used 
the $I_c - K$ color to derive a disk fraction of 61\%-88\% in the ONC,
and the Ca II lines in emission, or "filled-in", as a proxy for 
determining an accretion disk frequency of $\sim 70$\%.
{\it Rebull et al.} (2000) studied a region on both sides of the ONC
and determined a disk accretion fraction in excess of 40\%. 
{\it Lada et al.} (2000) used the JHKL bands
to derive a ONC disk fraction of 80\%-85\% in the low-mass PMS
population. {\it Lada et al.} (2004) extended this study to the substellar
candidate members and found a disk fraction of $\sim 50$\%.
In their ongoing large scale study of the Orion OB1 association, {\it Calvet et al.} (2005a)
combined UV, optical, JHKL and 10~$\mu$m measurements in a sample 
of confirmed members of the 1a and 1b sub-associations
to study dust emission and disk accretion. 
They showed evidence for an overall decrease in IR emission with age, 
interpreted as a sign of dust evolution between the disks in Ori OB 1b (age $\sim 4$ Myr),
Ori OB 1a (age $\sim 8$ Myr), and those of younger populations like Taurus (age $\sim 2$ Myr).
{\it Brice\~no et al.} (2005b) used IRAC and MIPS on Spitzer to look for 
dusty disks in Ori OB 1a and 1b. 
They confirm a decline in IR emission by the age of Orion OB 1b,
and find a number of "transition" disk systems ($\sim 14$\% in 1b and $\sim 6$\% in 1a), objects 
with essentially photospheric fluxes at wavelengths $\le 4.5 \> \mu$m and excess 
emission at longer wavelengths. These systems are interpreted as
showing signatures of inner disk clearing, with optically thin inner regions stretching
out to one or a few AU ({\it Calvet et al.}, 2002; {\it Uchida et al.}, 2004; {\it Calvet et al.}, 2005b;
{\it D'Alessio et al.}, 2005); the fraction of these transition disks that 
are still accreting is low ($\sim 5-10$\%), hinting at a
rapid shut off of the accretion phase in these systems
(similar results have been obtained by {\it Sicilia-Aguilar et al.}, 2006 in Cep OB2, see below).
{\it Haisch et al.} (2000) derived an IR-excess fraction of $\sim 86$\% in the 
$\simless 1$ Myr old NGC 2024 embedded cluster. Their findings indicate
that the majority of the sources that formed in NGC 2024 are presently
surrounded by, and were likely formed with, circumstellar disks.

One of the more surprising findings in the $\lambda$ Ori region was that,
despite the discovery of 72 low-mass PMS stars within $0.5 \arcdeg$ ($\sim 3.5$ pc)
of $\lambda$ Ori,
only two of them showed strong H$\alpha$ emission indicative of accretion disks
({\it Dolan and Mathieu}, 1999). {\it Dolan and Mathieu} (2001) expanded on this result by examining
the distribution of H$\alpha$ emission along an axis from B35 through $\lambda$ Ori to B30.
The paucity of H$\alpha$ emission-line stars continues from $\lambda$ Ori
out to the two dark clouds, at which point the surface density of H$\alpha$ emission-line
stars increases dramatically.
Even so, many of the H$\alpha$ stars associated with B30 and B35 have ages similar to
PMS stars found in the cluster near $\lambda$ Ori. Yet almost none of the latter show
H$\alpha$ emission. This strongly suggests that the absence of H$\alpha$ emission
from the central PMS stars is the result of an environmental influence linked
to the luminous OB stars.

The nearby Scorpius-Centaurus OB association (d $\sim 130$~pc, {\it de Zeeuw et al.}, 1999)
has been a natural place to search for circumstellar disks, especially at somewhat
older ages (up to $\sim 10$ Myr).
{\it Moneti et al.} (1999) detected excess emission at the ISOCAM
6.7 and $15\> \mu$m bands in 10 X-ray selected
WTTS belonging to the more widely spread population of Sco-Cen,
albeit at levels significantly lower than in the much 
younger (age $\sim 1$ Myr) Chamaeleon I T association.
In a sample of X-ray and
proper motion-selected late-type stars in the Lower Centaurus Crux (LCC, age $\sim 17$ Myr)
and Upper Centaurus Lupus (UCL, age $\sim 15$ Myr) in Sco-Cen,
{\it Mamajek et al.} (2002) find that
only 1 out of 110 PMS solar-type stars shows both enhanced H$\alpha$
emission and a K-band excess indicative of active accretion from a
truncated circumstellar disk, suggesting
time scales of $\sim 10$ Myr for halting most of the disk
accretion.
{\it Chen et al.} (2005) obtained Spitzer Space Telescope MIPS observations 
of 40 F- and G-type common proper motion members of the Sco-Cen OB association
with ages between 5 and 20 Myr. They detected $24\>\mu$m excess
emission in 14 objects, corresponding to a disk fraction of $\ge 35$\%.

\begin{figure}[ht]
\epsscale{1.0}
\plotone{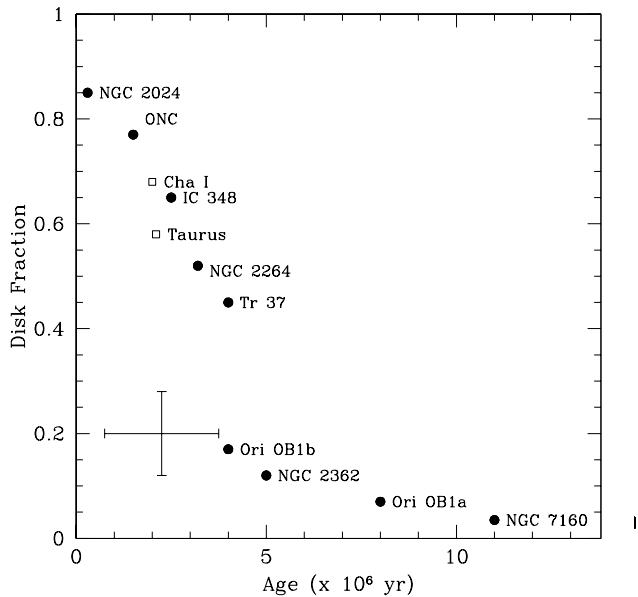}
\caption{\small
Inner disk fraction around low-mass young stars ($0.2 \la M \la 1~\msun$)
as a function of age in various nearby OB associations (solid dots).
The ONC point is an average from estimates by
{\it Hillenbrand et al.} (1998), {\it Lada et al.} (2000), {\it Haisch et al.} (2001).
Data in the JHKL bands from {\it Haisch et al.} (2001) were used for IC 348, NGC 2264, NGC 2362.
Note that for the more distant regions like NGC 2362 and NGC 2264, the mass 
completeness limit is $\sim 1~\msun$.
Results for Orion OB 1a and 1b are from JHK data in {\it Brice\~no et al.} (2005a).
In Tr 37 and NGC 7160 we used values by {\it Sicilia-Aguilar et al.} (2005, 2006)
from accretion indicators and JHK[3.6~$\mu$m] measurements.
In
As a comparison, we also plot disk fractions (open squares)
derived in Taurus (from data in {\it Kenyon and Hartmann}, 1995) and
for Chamaeleon I from {\it G\'omez and Kenyon} (2001).
A conservative errorbar is indicated at the lower left corner.
}
\label{disk_evol}
\end{figure}

In Perseus, {\it Haisch et al.} (2001) used JHKL observations to estimate a 
disk fraction of $\sim 65$\% in the 2-3 Myr old ({\it Luhman et al.}, 1998) IC 348 cluster,
a value lower than in the younger NGC 2024 and ONC clusters in Orion,
suggestive of a timescale of ~2-3 Myr for the disappearance of $\sim 1/3$ of
the inner disks in IC 348. 
The $\sim 4$ Myr old Tr 37 and the $\sim 10$ Myr old NGC 7160 clusters
in  Cep OB2 have been studied by {\it Sicilia-Aguilar et al.} (2005),
who derive an accreting fraction of
$\sim 40$\% for Tr 37, and 2-5\% (1 object) for NGC 7160.
{\it Sicilia-Aguilar et al.} (2006) used Spitzer IRAC and MIPS observations to
further investigate disk properties in Cep OB2. About 48\% of the members
exhibit IR excesses in the IRAC bands, consistent with their inferred accreting
disk fraction. They also find a number of "transition" objects (10\%) in Tr 37.
They interpret their results as evidence for differential evolution
in optically thick disks as a function of age, with faster decreases in the IR emission
at shorter wavelengths, suggestive of a more rapid evolution in the inner disks.

All these studies agree on a general trend towards rapid inner disk
evolution (Fig.~5); it looks like the disappearance of the
dust in the innermost parts of the disk, either due to grain growth and settling or to
photo-evaporation ({\it Clarke et al.}, 2001), is followed by a rapid shut off of accretion.
However, both these investigations and newer 
findings also reveal exceptions that lead to a more complex picture.
{\it Muzerolle et al.} (2005) find that 5-10\% of all disk sources in the $\sim 1$ Myr old
NGC 2068 and 2071 clusters in Orion show evidence of significant grain growth, 
suggesting a wide variation in timescales for the onset of 
primordial disk evolution and dissipation.  This could also be related to the 
existence of CTTS and WTTS in many regions at very young ages, the WTTS showing 
no signatures of inner, optically thick disks. Did some particular initial conditions
favor very fast disk evolution in these WTTS (binarity)?
Finally, the detection of long-lived disks implies that dust dissipation
and the halting of accretion do not necessarily follow a universal trend.
This may have implications for the formation of planetary systems;
if slow accretion processes are the dominant formation mechanism for Jovian planets
then long-lived disks may be ideal sites to search for evidence for protoplanets.

\bigskip
\noindent
\section{\textbf{THE ORIGIN OF OB ASSOCIATIONS:\\
BOUND VS. UNBOUND STELLAR GROUPS}}
\bigskip

{\it Blauuw} (1964) in his masterly review first provided 
some clues as to the origin of OB associations,
alluding to two ideas: "originally small, compact bodies with
dimensions of several parsec or less" or "regional star
formation, more or less simultaneous, but scattered over
different parts of a large cloud complex". 
Similarly, there are at present two (competing) models that
attempt to explain the origin of OB associations:

A) origin as expanding dense embedded clusters

B) origin in unbound turbulent giant molecular clouds

We will label Model A as the KLL-model
({\it Kroupa et al.}, 2001; {\it Lada and Lada}, 2003),
while Model B will be named the BBC-model (for Bonnell, Bate and
Clark,  see also {\it Clark et al.}, 2005).

The jury is still out on which model fits the observations better.
It may be that both models contain elements of truth and thus
are not mutually exclusive. Future astrometric surveys, $Gaia$
in particular, will provide constraints to perform sensitive
tests on these models,  but that will not happen for about another decade.
We refer the reader to the review of {\it Brown et al.} (1999), who also points out
the problems involved in the definition of an OB association and 
the division of these vast stellar aggregates 
into subgroups (see {\it Brown et al.}, 1997).

We now discuss both models in turn.
The KLL model proposes that ultimately the star formation
efficiency in most embedded, incipient star clusters is too low
($\simless 5$ percent) for these clusters to remain bound (see references
in {\it Elmegreen et al.}, 2000; {\it Lada and Lada}, 2003) after
the massive stars have expelled the bulk of the lower density,
left-over cluster gas that was not turned into stars, or
did not get accreted. Therefore the cluster will find
itself globally unbound, although the cluster {\it core} may
remain bound ({\it Kroupa et al.}, 2001) and survive.
Such an originally compact but expanding cluster could evolve
into an extended subgroup of an OB association after a few
Myr. 
Essentially, the KLL model describes how the
dense gas in a bound and virialized GMC is partly converted
into stars, most of which are born in dense embedded clusters,
the majority of which disperse quickly (so-called cluster
"infant mortality").

The BBC model supposes that GMCs need not be regarded as
objects in virial equilibrium, or even bound, for them to
be sites of star formation. Globally unbound or marginally
bound GMC can form stellar groups or clusters very quickly,
over roughly their crossing time (cf. {\it Elmegreen et al.}, 2000; 
recent simulations also show that GMC themselves
can form very quickly from atomic gas, in 1-2 Myr
(see chapter by {\it Ballesteros-Paredes et al.}).
The unbound state of the GMC ensures that the
whole region is dispersing while it is forming stars or
star clusters in locally compressed sheets and filaments,
due to the compressive nature of supersonic turbulence
("converging flows"). The mass fraction of compressed cloud gas
is low, affecting only $\sim 10$\% of the GMC.
In this model, the spacing of the OB stars that define an
OB association would be initially large (larger than a few pc),
rather than compact ($\la 1$ pc) as in the KLL model.
Furthermore, no sequential star formation triggered by the thermal
pressure of expanding HII regions "a la {\it Elmegreen and Lada} (1977)"
may be needed to create OB subgroups, unlike in the KLL model.
Delayed SN-triggered star formation in adjacent transient clouds
at some distance may still occur and add to the geometric
complexity of the spatial and temporal distribution of stars
in giant OB associations (e.g., NGC 206 in M31 and NGC 604 in M33;
{\it Ma{\'\i}z-Apell\'aniz et al.}, 2004).

A problem common to both models is that neither one explains
very well the spatial and temporal structure of OB associations,
i.e. the fact that the subgroups seem to form an age sequence
(well known in Orion OB1 and Sco OB2). One way out would be to
argue that the subgroups are not always causally connected
and did not originate by sequential star formation as in the
{\it Elmegreen and Lada} (1977) paradigm. Star formation in clouds
with supersonic turbulence occurring in convergent flows, may
be of a more random nature and only mimic a causal sequence
of triggering events (e.g., in Sco OB2, the UCL and LCC subgroups,
with ages of $\sim 15$ and $\sim 17$ Myr from the 
{\it D'Antona and Mazzitelli} [1994] tracks, 
have hardly an age difference at all).
This is a problem for the KLL model, as sequential star formation
can hardly generate two adjacent clusters (subgroups) within such
a short time span. It is also a problem for the BBC model, but
for a different reason. The fact that star formation must be
rapid in unbound transient molecular clouds is in conflict with
the ages of the Sco OB2 subgroups (1, 5, 15 and 17 Myr),
if all subgroups formed from a single coherent (long-lived) GMC.
Still, the observational evidence suggesting that Upper Sco can be
understood in the context of triggered star formation
(see Section~7.2), can be reconciled with these models if we consider 
a scenario with multiple star formation sites in a turbulent large
GMC, in which triggering may easily take place.
It remains to be seen if subgroups of OB associations had some
elongated minimum size configuration, as {\it Blaauw} (1991) surmised,
of the order of 20 pc x 40 pc.
If so, OB associations
are then something fundamentally different from embedded clusters,
which would have many ramifications for the origin of OB stars
(However, a caveat with tracing back minimum size configurations [{\it Brown et al.}, 1997]
is that, even using modern {\it Hipparcos} proper motions,
there is a tendency to obtain overestimated dimensions
because present proper motions cannot resolve the
small velocity dispersion. This situation should improve
with $Gaia$, and with the newer census of low-mass stars, that can
potentially provide
statistically robust samples to trace the past kinematics of
these regions).

\bigskip
\noindent
\section{\textbf{CONSTRAINTS ON RAPID AND \\
TRIGGERED/SEQUENTIAL STAR FORMATION}}
\bigskip

\textbf{ 7.1 The duration of star formation}
\medskip

One of the problems directly related to the properties of the stellar
populations in OB associations are the lifetimes
of molecular clouds. Age estimates for GMCs can be very discordant, ranging
from $\sim 10^8$ years ({\it Solomon et al.}, 1979; 
{\it Scoville and Hersh}, 1979) to just a few $10^6$ years (e.g., {\it Elmegreen et al.},
 2000; {\it Hartmann et al.}, 2001; {\it Clark et al.}, 2005).
Molecular clouds lifetimes bear importantly on the
picture we have of the process of star formation. Two main views have been
contending among the scientific community during the past few years.
In the standard picture of star formation magnetic fields
are a major support mechanism for clouds ({\it Shu et al.}, 1987). 
Because of this, the cloud must
somehow reduce its magnetic flux per unit mass if it is to attain the
critical value for collapse. One way to do this is through ambipolar
diffusion, in which the gravitational force pulls mass through the resisting
magnetic field, effectively concentrating the cloud and slowly
"leaving the magnetic field behind".
>From these arguments it follows that timescale for
star formation should be of the order of the diffusion time of the magnetic
field, $t_D \sim 5 \times 10^{13} (n_i / n_{H2})$ yr ({\it Hartmann}, 1998), which will
be important only if the ionization inside the cloud is low
($n_i/n_{H2} \la 10^{-7}$), in which case $t_D \sim 10^7$ yr; therefore, the so
called "standard" picture depicts star formation as a "slow" process.
This leads to relevant observational consequences.
If molecular clouds live for long periods before
the onset of star formation, then
we should expect to find a majority of starless dark clouds;
however, the observational evidence points to quite the contrary.
Almost all cloud complexes within $\sim 500$ pc
exhibit active star formation, harboring 
stellar populations with ages $\sim 1-10$ Myr.
Another implication of "slow" star formation is that of {\it age
spreads} in star-forming regions.  If clouds such as Taurus last
for tens of Myr there should exist a population of PMS
sequence stars with comparable ages (the "post-T Tauri problem").
Many searches for such "missing population"
were conducted in the optical and in X-rays, in
Taurus and in other regions (see {\it Neuh\"auser} 1997).
The early claims by these studies of the detection
of large numbers of older T Tauri stars widely spread
across several nearby star forming regions, were countered by
{\it Brice\~no et al.} (1997), who showed that these samples were composed of an
admixture of young, x-ray active ZAMS field stars and some true PMS
sequence members of these regions. Subsequent high-resolution spectroscopy
confirmed this idea.
As discussed in Section~4, presently 
there is little evidence for the presence of substantial numbers
of older PMS stars in and around molecular clouds.

Recent wide-field optical studies in OB associations like Sco-Cen ({\it Preibisch et al.}, 2002),
Orion ({\it Dolan and Mathieu}, 2001; {\it Brice\~no et al.}, 2001, 2005, 2006), and Cepheus
({\it Sicilia-Aguilar et al.}, 2005), show that the groupings of stars with ages $\ga
4-5$ Myr have mostly lost their natal gas.
The growing notion is that not only do molecular clouds form stars rapidly,
but that they are transient structures, dissipating quickly after the
onset of star formation. This dispersal seems to be effective in both
low-mass regions as well as in GMC complexes that give birth to OB associations.
The problem of accumulating and then dissipating the gas quickly in
molecular clouds has been addressed by {\it Hartmann et al.} (2001).
The energy input from stellar winds of massive stars,
or more easily from SN shocks, seems to be able to
account for the dispersal of the gas on short timescales in the high density
regions typical of GMC complexes, as well as in low-density
regions, like Taurus or Lupus;
if enough stellar energy is input into the gas such that
the column density is reduced by factors of only 2-3,
the shielding could be reduced enough to allow dissociation of much
of the gas into atomic phase, effectively "dissipating" the molecular cloud.

\medskip
\textbf{ 7.2 Sequential and triggered star formation}
\medskip

{\it Preibisch et al.} (2002) investigated the star formation history 
in Upper Scorpius.
A very important aspect in this context is the spatial extent of the
association and the corresponding crossing time.
The bulk (70\%) of the {\it Hipparcos} members (and thus also the
low-mass stars) lie within an area of 11 degrees diameter on the sky,
which implies a characteristic size of the association of 28~pc.
They estimated that the original size of the association 
was probably about 25~pc.
{\it de Bruijne} (1999) showed that the internal velocity dispersion of the {\it Hipparcos}
members of Upper Sco is only 1.3 $\rm km \> s^{-1}$.
This implies a lateral
crossing time of 25~pc~/~1.3 $\rm km \> s^{-1}$ $\sim 20$~Myr.
It is obvious that the lateral crossing time is much
(about an order of magnitude) larger than the age spread of the
association members (which is $< 2$~Myr as derived by {\it Preibisch and Zinnecker},
1999).  This finding clearly shows that some external agent is
required to have coordinated the onset of the star formation process over the
full spatial extent of the association. In order to account for the small
spread of stellar ages, the triggering agent must have crossed the initial
cloud with a velocity of at least $\rm \sim 15\!-\!25 \> km \> s^{-1}$.
Finally, some mechanism must have terminated the star formation process at
most about 1 Myr after it started.  Both effects can be attributed to
the influence of massive stars.

In their immediate surroundings, massive stars generally have a
 destructive effect on their
environment; they can disrupt molecular clouds very quickly and therefore
prevent further star formation.  At somewhat larger
distances, however, the wind- and supernova-driven shock waves originating
from massive stars can have a constructive rather than destructive effect by
driving molecular cloud cores into collapse.  Several numerical studies
(e.g., {\it Boss}, 1995; {\it Foster and Boss}, 1996; 
{\it Vanhala and Cameron}, 1998; {\it Fukuda and Hanawa}, 2000) 
have found that the outcome of the impact of a shock wave on a
cloud core mainly depends on the type of the shock and its velocity: In its
initial, adiabatic phase, the shock wave is likely to destroy ambient clouds;
the later, isothermal phase, however, is capable of triggering cloud collapse
if the velocity is in the right range.  Shocks traveling faster than about 50 $\rm \> km\> s^{-1}$
shred cloud cores to pieces, while shocks with velocities slower than
about 15 $\rm \> km\> s^{-1}$ usually cause only a slight temporary compression of cloud
cores.  Shock waves with velocities in the range of $\sim 15 - 45$ $\rm \> km\> s^{-1}$,
however, are able to induce collapse of molecular cloud cores.
A good source of shock waves with velocities in that range are supernova
explosions at a distance between $\sim 10$ pc and $\sim 100$ pc.
Other potential sources of such shock waves include wind-blown bubbles and
expanding HII regions.  Observational evidence for star forming events
triggered by shock waves from massive stars has for example been discussed in
{\it Carpenter et al.} (2000), {\it Walborn et al.} (1999), {\it Yamaguchi et al.} (2001),
{\it Efremov and Elmegreen} (1998), {\it Oey and Massey} (1995), and {\it Oey et al.} (2005).

For the star burst in Upper Sco, a very suitable trigger is a supernova
explosion in the Upper Centaurus-Lupus association that happened about 12 Myr
ago.  The structure and kinematics of the large H I loops surrounding the
Scorpius-Centaurus association suggest that this shock wave passed through the
former Upper Sco molecular cloud just about 5--6 Myr ago ({\it de Geus}, 1992).  This
point in time agrees very well with the ages found for the low-mass stars as
well as the high-mass stars in Upper Sco, which have been determined above in
an absolutely independent way.  Furthermore, since the distance from Upper
Centaurus-Lupus to Upper Sco is about 60 pc, this shock wave probably had
precisely the properties ($v \sim 20\!-\!25$ $\rm \> km\> s^{-1}$) that are required to induce
star formation according to the modeling results mentioned above.  Thus, the
assumption that this supernova shock wave triggered the star formation process
in Upper Sco provides a self-consistent explanation of all observational data.

The shock-wave crossing Upper Sco initiated the formation of some 2500 stars,
including 10 massive stars upwards of $10\,M_\odot$.
When the new-born massive stars
`turned on', they immediately started to destroy the cloud from inside by
their ionizing radiation and their strong winds. This affected the cloud so
strongly that after a period of $\la 1$ Myr the star formation process
was terminated, probably simply because all the remaining dense cloud material
was disrupted. This explains the narrow age distribution and why only
about 2\% of the original cloud mass was transformed into stars.  
About 1.5 Myr ago the most massive star in Upper Sco, probably
the progenitor of the pulsar PSR J1932+1059, exploded as a supernova.
This explosion created a strong shock wave, which fully dispersed the Upper Sco 
molecular cloud and removed basically all the remaining diffuse material.

It is interesting to note that this shock wave must have
crossed the $\rho$ Oph cloud within the last 1 Myr ({\it de Geus}, 1992).
The strong star formation
activity we witness right now in the $\rho$ Oph cloud
might therefore be triggered by this shock wave (see {\it Motte et al.}, 1998) and
would represent the third generation of sequential triggered
star formation in the Scorpius-Centaurus-Ophiuchus complex.

Other relatively nearby
regions have also been suggested as scenarios for triggered
star formation. 
In Cepheus, a large scale ring-like feature with a diameter of
120 pc has been known since the time
of the H$\alpha$ photographic atlases of HII regions ({\it Sivan}, 1974).
{\it Kun et al.}, (1987) first identified the infrared emission of
this structure in $IRAS$ 60 and 100~$\mu$m sky flux maps.
The Cepheus bubble includes the Cepheus OB2 association (Cep OB2), 
which is partly made up of the Tr 37 and NGC 7160 open clusters,
and includes the HII region IC 1396.
{\it Patel et al.} (1995, 1998) mapped $\rm \sim 100\> deg^2$ in Cepheus in the
$J = 1 - 0$ transition of CO and $\rm ^{13}CO$. Their observations reveal
that the molecular clouds are undergoing an asymmetrical expansion away
from the Galactic plane.
They propose a scenario in which 
the large scale bubble was blown away by stellar winds and photoionization from
the first generation of OB stars, which are no longer present (having exploded
as supernovae). The $\sim 10$ Myr old ({\it Sicilia-Aguilar et al.}, 2004) 
NGC 7160 cluster and evolved stars such as $\mu$ Cephei, VV Cephei and $\nu$ Cephei
are the present day companions of those first OB stars.
{\it Patel et al.} (1998) show that the expanding shell becomes unstable at $\sim 7$ Myr
after the birth of the first OB stars. The estimated radius of the shell at that
time ($\sim 30$ pc) is consistent with the present radius of the ring of O and B-type
stars which constitute Cep OB2. Within a factor of $\la 2$, this age is also consistent with 
the estimated age for the Tr 37 cluster ($\sim 4$ Myr; {\it Sicilia-Aguilar et al.}, 2004).
Once the second generation of massive stars formed, they started affecting the dense gas
in the remaining parent shell. The gas around these O stars expanded in rings like the
one seen in IC 1396. The dynamical timescale for this expansion is of the order of 1-3 Myr,
consistent with the very young ages ($\sim 1-2$ Myr) of the low-mass stars 
in the vicinity of IC 1396 ({\it Sicilia-Aguilar et al.}, 2004). This HII region is
interpreted as the most recent generation of stars in Cep OB2.

In the Orion OB1 association
{\it Blaauw} (1964) proposed that the 
ONC is the most recent event in a series of star-forming 
episodes within this association. The increasing ages between
the ONC, Ori OB1b and Ori OB1a have been suggested to be a case for
sequential star formation ({\it Blaauw}, 1991).
However, until now it has been difficult to investigate triggered
star formation in Ori OB1 because of the lack of an unbiased
census of the low-mass stars over the entire region.
This situation is changing with the new large scale surveys 
(e.g., {\it Brice\~no et al.}, 2005) that are
are mapping the low-mass population of Ori OB1 over tens of square
degrees; we may soon be able to test if Orion
can also be interpreted as a case of induced, 
sequential star formation.

\bigskip
\noindent
\section{\textbf{CONCLUDING REMARKS}}
\bigskip

Low-mass stars ($\rm 0.1 \la M \la 1~\msun$)
in OB associations are essential for understanding many of
the most fundamental problems in star formation, and important progress has been
made during the past years by mapping and characterizing these objects.
The newer large scale surveys reveal that low-mass stars
exist wherever high mass stars are found, not only in the dense clusters, but also
in a much more widely distributed population. As ever increasing numbers of 
low-mass stars are identified over large areas in older regions
like Orion OB1a, their spatial distribution shows substructure suggestive of a far
more complex history than would be inferred from the massive stars.

The low-mass stellar populations in OB associations provide a snapshot of
the IMF just after the completion of star formation, and before stars
diffuse into the field population.
The IMF derived from OB associations is consistent with the field IMF.
The large majority of the low-mass PMS stars in the solar vicinity are in
OB associations, therefore this agrees with early suggestions ({\it Miller and Scalo}, 1978)
that the majority of stars in the Galaxy were born in OB associations.

Since PPIV, the recent large surveys for low-mass members in several 
OB associations have allowed important progress on studies of early circumstellar
disk evolution.
With large scale surveys like 2MASS, large IR imagers and now the Spitzer Space Telescope,
we have unprecedented amounts of data sensitive to dusty disks in many regions.
Overall disks largely dissipate over timescales of a few Myr, either by dust evaporation,
or settling and growth into larger bodies like planetesimals and planets; exactly which
mechanisms participate in this evolution may depend on initial conditions and even on
the environment. The actual picture seems more complex, current evidence supports 
a wide range of disk properties even at ages of $\sim 1$ Myr, and in some regions 
disks somehow manage to extend their lifetimes, surviving for up to $\sim 10-20$ Myr.

As the census of low-mass stars in nearby OB associations are extended 
in the coming years, our overall picture of
star formation promises to grow even more complex and challenging.

\vspace{0.2cm}

\textbf{ Acknowledgments.} 

We thank the referee, Kevin Luhman, for his thorough review and
useful suggestions that helped us improve this manuscript. We also
are grateful to A.~G.~A Brown for helpful comments. 
C. Brice\~no acknowledges support from NASA Origins grant NGC-5 10545.
E. Mamajek is supported through a Clay Postdoctoral Fellowship from the
Smithsonian Astrophysical Observatory.
R. Mathieu appreciates the support of the National Science Foundation.
F. Walter acknowledges support from NSF grant AST-030745 to Stony Brook University.
T. Preibisch and H. Zinnecker are grateful to the 
Deutsche Forschungsgmeinschaft for travel support to attend PPV.

\bigskip

\centerline\textbf{ REFERENCES}
\bigskip
\parskip=0pt
{\small
\baselineskip=11pt

\refs Allen L.~E., Calvet N., D'Alessio P., Merin B., Hartmann L., et al.  (2004) \apjs, {\it 154}, 363-366.

\refs Ambartsumian V.~A. (1947) In {\it Stellar Evolution and Astrophysics, 
Armenian Acad. of Sci.} (German translation, 1951, Abhandl, Sowjetischen Astron., 1, 33.).

\refs Ballesteros-Paredes J., Hartmann L., and V\'azquez-Semadeni E. (1999)  \apj, {\it 527}, 285-297.

\refs Baraffe I., Chabrier G., Allard F., and Hauschildt P.H. (1998) \aap, {\it 337}, 403-412.

\refs Baraffe I., Chabrier G., Allard F., and Hauschildt P.H. (2001) In
IAU Symp. 200 {\it The Formation of Binary Stars,} (Zinnecker H. and
Mathieu R.~D., eds.), pp. 483-491. ASP, San Francisco.

\refs Barrado y Navascu\'es D., Stauffer J.~R., Bouvier J., Jayawardhana R., and Cuillandre J-C. (2004) \apj, {\it 610}, 1064-1078.

\refs Beckwith S.~V.~W., Sargent A.~I., Chini R.~S., and Guesten, R.  (1990) \aj, {\it 99}, 924-945.

\refs Bertelli G., Bressan A., Chiosi C., Fagotto F., and Nasi E., 1994, \aaps, {\it 106}, 275-302.

\refs Blaauw A. (1964) \araa, {\it 2}, 213-246.	

\refs Blaauw A. (1991) In {\it The Physics of Star Formation and Early Stellar Evolution, 
NATO Advanced Science Institutes (ASI) Series C} 
(C.~J. Lada and N.~D. Kylafis, eds.), {\it 342}, pp.125-154. Kluwer, Dordrecht.

\refs Boss, A.~P. (1995) \apj, {\it 439}, 224-236.

\refs Brandl B., Brandner W., Eisenhauer F., Moffat A.~F.~J., Palla F.,
and Zinnecker H. (1999) \aap, {\it 352}, L69-L72.

\refs Brandner W., Grebel E.~K., Barb\'a R.~H., Walborn N.~R., and Moneti A. (2001) \aj, {\it 122}, 858-865.

\refs Brice\~no C., Calvet N., G\'omez M., Hartmann L., Kenyon S., and Whitney B. (1993) \pasp, {\it 105}, 686-692.

\refs Brice\~no C., Hartmann L.~W., Stauffer J., Gagn\'e M., Stern R., and Caillault J. (1997) \aj, {\it 113}, 740-752.

\refs Brice\~no C., Hartmann L., Calvet N., and Kenyon, S. (1999) \aj, {\it 118}, 1354-1368.

\refs Brice\~no C., Vivas A.~K., Calvet N., Hartmann L. et al. (2001) \science, {\it 291}, 93-96.

\refs Brice\~no C., Calvet N., Hern\'andez J., Vivas A.~K., Hartmann L., Downes J.~J., and Berlind P. (2005a) \aj, {\it 129}, 907-926.

\refs Brice\~no C., Calvet N., Hern\'andez J., Hartmann L., Muzerolle J., D'Alessio P., and Vivas A.~K. (2005b) In 
{\it Star Formation in the Era of Three Great Observatories}, http://cxc.harvard.edu/stars05/agenda/program.html.

\refs Brice\~no C., Calvet N., Hern\'andez J., Vivas A.~K., Hartmann L., Downes J.~J., and Berlind P. (2006) \apj, submitted

\refs Brown A.~G.~A. (1996) \pasp, {\it 108}, 459-459.

\refs Brown A.~G.~A., de Geus E.~J., de Zeeuw P.~T. (1994), \aap, {\it 289}, 101-120.

\refs Brown A.~G.~A., Dekker G., and de Zeeuw P.~T. (1997) \mnras, {\it 285}, 479-492.

\refs Brown A.~G.~A., Blaauw A., Hoogerwerf R., de Bruijne J.~H.~J., and de Zeeuw P.~T. (1999) 
In {\it The Origin of Stars and Planetary Systems} (C.~J. Lada and N.~D. Kylafis, eds.), pp. 411-440.
Kluwer, Dordrecht.

\refs Burningham B., Naylor T., Littlefair S.~P., and Jeffries R.~D. (2005) \mnras, {\it 363}, 1389-1397.

\refs Calvet N. and Gullbring E. (1998) \apj, {\it 509}, 802-818.

\refs Calvet N., D'Alessio P., Hartmann L., Wilner D., Walsh A., and Sitko M. (2002) \apj, {\it 568}, 1008-1016.

\refs Calvet N., Brice\~no C., Hern\'andez J., Hoyer S., Hartmann L., et al. (2005a) \aj, {\it 129}, 935-946.

\refs Calvet N., D'Alessio P., Watson D.~M., Franco-Hern\'andez R., Furlan E.  et al.  (2005b) \apj, {\it 630}, L185-L188.

\refs Carpenter J.~M., Heyer M.~H., and Snell R.~L. (2000) \apjs, {\it 130}, 381-402.

\refs Carpenter J.~M., Hillenbrand L.~A., and Strutskie M.~F. R. (2001) \aj, {\it 121}, 3160-3190.

\refs Chen C.~H., Jura M., Gordon K.~D., and Blaylock M. (2005) \apj, {\it 623}, 493-501.

\refs Clark P.~C., Bonnell I.~A., Zinnecker H., and Bate M.~R. (2005) \mnras, {\it 359}, 809-818.

\refs Clarke C.~J., Gendrin A., and Sotomayor, M. (2001) \mnras, {\it 328}, 485-491.

\refs D'Alessio P., Hartmann L., Calvet N., Franco-Hern\'andez R., Forrest W. et al. (2005) \apj, {\it 621}, 461-472.

\refs D'Antona F. and Mazzitelli I. (1994) \apjs, {\it 90}, 467-500.

\refs de Bruijne J.~H.~J. (1999) \mnras, {\it 310}, 585-617.

\refs de Geus, E.~J., 1992, \aap, {\it 262}, 258-270.

\refs de Geus E.~J., de Zeeuw P.~T., and Lub J. (1989) \aap {\it 216}, 44-61.

\refs de Zeeuw  P.~T. and  Brand J. (1985) In {\it 
Birth and Evolution of Massive Stars and Stellar Groups}
(van den Woerden H. and Boland W., eds.), pp. 95-101.  Reidel, Dordrecht.

\refs de Zeeuw P.~T., Hoogerwerf R., de Bruijne J.~H.~J., Brown A.~G.~A., and Blaauw A. (1999) \aj, {\it 117}, 354-399.

\refs Dolan Ch.~J. and Mathieu R.~D. (1999) \aj, {\it 118}, 2409-2423.

\refs Dolan Ch.~J. and Mathieu R.~D. (2001) \aj, {\it 121}, 2124-2147.

\refs Efremov Y. and Elmegreen, B.~G. (1998) \mnras, {\it 299}, 643-652.

\refs Elmegreen B.~G. (1990) In {\it The evolution of the interstellar medium} (Blitz, L. ed.),  pp. 247-271. ASP, San Francisco.

\refs Elmegreen B.~G. and Lada C.~J. (1977) \apj, {\it 214},  725-741.

\refs Elmegreen B.~G., Efremov Y., Pudritz R.~E., and Zinnecker H. (2000) In
{\it Protostars and Planets IV} (Mannings V., Boss A.P., and Russell S. S., eds.),
pp. 179-202.  Univ. of Arizona, Tucson.

\refs Favata F. and  Micela G. (2003) {\it Space Science Reviews, 108}, 577-708.

\refs Feigelson E.~D. and DeCampli W.~M. (1981) \apj, {\it 243}, L89-L93.

\refs Feigelson E.~D. and Montmerle T. (1999) \araa, {\it 37}, 363-408.

\refs Foster P.~N. and Boss A.~P. (1996) \apj, {\it 468}, 784-796.

\refs Fukuda N. and Hanawa T. (2000) \apj, {\it 533}, 911-923.

\refs Getman K.~V., Flaccomio E., Broos P.~S., Grosso N., Tsujimoto M. et al. (2005) \apjs, {\it 160}, 319-352.

\refs G\'omez M. and Kenyon S.~J. (2001) \aj, {\it 121}, 974-983.

\refs G\'omez M. and Lada C. (1998) \aj, {\it 115}, 1524-1535.

\refs Gullbring E., Hartmann L., Brice\~no C., and Calvet N. (1998) \apj, {\it 492}, 323-341.

\refs Haisch K.~E. Jr., Lada E.~A., and Lada C.~J. (2000) \aj, {\it 120}, 1396-1409.

\refs Haisch K.~E. Jr., Lada E.~A., and Lada C.~J. (2001) \apj, {\it 553}, L153-L156.


\refs Hartmann L. (1998) In {Accretion processes in star formation, Cambridge astrophysics series}, 
Cambridge University Press, Cambridge.

\refs Hartmann L. (2001) \aj, {\it 121}, 1030-1039.

\refs Hartmann L., Stauffer J.~R., Kenyon S.~J., and Jones B.~F.  (1991) \aj, {\it 101}, 1050-1062.

\refs Hartmann L., Ballesteros-Paredes J., and Bergin E.~A. (2001) {\it 562}, 852-868.

\refs Hartmann L., Calvet N.,  Watson D.M., D'Alessio P., Furlan E., et al.  (2005a) \apj, {\it 628}, L147-L150.

\refs Hartmann L., Megeath S.~T., Allen L., Luhman K., Calvet N. et al. (2005b) \apj, {\it 629}, 881-896.

\refs Herbig G.~H. (1962) {\it Adv. Astr. Astrophys., 1}, 47-103.

\refs Herbig G.~H. (1978) In {\it Problems of Physics and Evolution of the Universe.}
(L.V. Mirzoyan, ed.), pp.171-188. Pub. Armenian Academy of Sciences, Yerevan.

\refs Herbig G.~H. and Bell K.~R. (1988) {\it Lick Observatory Bulletin}, Lick Observ., Santa Cruz.

\refs Herbst W., Herbst D.~K., Grossman E.~J., and Weinstein D. (1994) \aj, {\it 108}, 1906-1923.

\refs Hillenbrand L.~A. (1997) \aj, {\it 113}, 1733-1768.

\refs Hillenbrand L.~A., Strom S.~E., Calvet N., Merrill K.~M., Gatley I. et al. (1998) \aj, {\it 116}, 1816-1841.

\refs Hoogerwerf R. (2000) \mnras, {\it 313}, 43-65.


\refs Johns-Krull C.~M. and Valenti J.~A. (2001) \apj, {\it 561}, 1060-1073.

\refs Joy A.~H. (1945) \apj, {\it 102}, 168-200.


\refs Kenyon S.~J. and Hartmann L.~W. (1995) \apjs, {\it 101}, 117-171.

\refs Kenyon M.~J., Jeffries R.~D., Naylor T., Oliveira J.~M., and Maxted P.~F.~L. (2005) \mnras, {\it 356}, 89-106.

\refs Kholopov P.~N. (1959) {\it Sov. Astron., 3}, 425-433.

\refs K\"onigl A. (1991) \apj, {\it 370}, L39-L43.

\refs Kroupa P. (2002) \science, {\it 295}, 82-91.

\refs Kroupa P., Aarseth S., and Hurley J.  (2001) \mnras, {\it 321}, 699-712.

\refs Kun M., Balazs L.~G., and Toth I. (1987) {\it Astrophys. and Space Sci., 134}, 211-217.

\refs Lada C.~J. and Lada, E.~A. (2003) \araa, {\it 41}, 57-115.

\refs Lada C.~J., Muench A.~A., Haisch K.~E.~Jr., Lada E.~A., Alves J.~F. et al. (2000) \aj, {\it 120}, 3162-3176.

\refs Lada C.~J., Muench A.~A., Lada E.~A., and Alves J.~F. (2004) \aj, {\it 128}, 1254-1264.

\refs Larson R. (1985) \mnras, {\it 214}, 379-398.

\refs Lee H-T., Chen W.~P., Zhang Z-W., and Hu J-Y. (2005) \apj, {\it 624}, 808-820.

\refs Leitherer C. (1998) In {\it The Stellar Initial Mass Function (38th Herstmonceux Conference)}
(G. Gilmore and D. Howell, eds.). {\it 142}, pp. 61-88. ASP, San Francisco.

\refs Liu C.~P., Zhang, C.~S., and Kimura H. (1981) {\it Chinese Astron. Astrophys,  5}, 276-281.


\refs Luhman K. L. (1999) \apj, {\it 525}, 466-481.

\refs Luhman K.~L., Rieke G.~H., Lada C.~J., and Lada E.~A. (1998) \apj, {\it 508}, 347-369.


\refs Ma{\'\i}z-Apell\'aniz J., P\'erez E., and Mas-Hesse J.~M. (2004) \aj, {\it 128}, 1196-1218.

\refs Mamajek E.~E., Meyer M.~R., and Liebert J.~W. (2002) BAAS, {\it 34}, 762-762.

\refs Mart{\'\i}n E.~L., Rebolo R., and Zapatero-Osorio M.~R. (1996) \apj, {\it 469}, 706-714.

\refs McGehee P.~M., West A.~A., Smith J.~A., Anderson, K.~S.~J., and Brinkmann J. (2005) \aj, {\it 130}, 1752-1762.

\refs Meyer M., Calvet N., and Hillenbrand L.~A. (1997) \aj, {\it 114}, 288-300.

\refs Mikami T. and Ogura K.  (2001) \apss, {\it 275}, 441-462.

\refs Miller G.~E. and Scalo J.~M. (1978) \pasp, {\it 90}, 506-513.

\refs Moneti A., et al. (1999) In {\it Astrophysics with Infrared
  Surveys: A Prelude to SIRTF}, (M.~D. Bicay,
R.~M. Cutri, and B.~F. Madore, eds.), pp. 355-358. ASP, San Francisco.

\refs Motte F., Andr\'e P., and Neri R. (1998) \aap, {\it 336}, 150-172.

\refs Muzerolle J., Calvet N., and Hartmann L. (1998a) \apj, {\it 492}, 743-753.

\refs Muzerolle J., Hartmann L., and Calvet N. (1998b) \aj, {\it 116}, 455-468.

\refs Muzerolle J., Calvet N., and Hartmann L. (2001) \apj, {\it 550}, 944-961.

\refs Muzerolle J., Young E., Megeath S.~T., and Allen L. (2005) In {\it Star Formation in the Era of Three 
Great Observatories}, http://cxc.harvard.edu/stars05/agenda/program.html.

\refs Neuh\"auser R. (1997) \science, {\it 267}, 1363-1370.

\refs Oey M.~S. and Massey P. (1995) \apj, {\it 542}, 210-225.

\refs Oey M.~S., Watson A.~M., Kern K., and Walth G.~L. (2005) \aj, {\it 129}, 393-401.

\refs Ogura K. (1984) \pasj, {\it 36}, 139-148.


\refs Palla F. and Stahler S.~W. (1992) \apj, {\it 392}, 667-677.

\refs Palla F. and Stahler S.~W. (1993) \apj, {\it 418}, 414-425.

\refs Palla F. and Stahler S.~W. (1999) \apj, {\it 525}, 772-783.

\refs Palla F. and Stahler S.~W. (2000) \apj, {\it 540}, 255-270.

\refs Patel N.~A., Goldsmith P.~F., Snell R.~L., Hezel T., and Xie T. (1995) \apj, {\it 447}, 721-741.

\refs Patel N.~A., Goldsmith P.~F., Heyer M.~H., Snell R.~L., and Pratap P. (1998) \apj, {\it 507}, 241-253. 

\refs Preibisch Th. and  Zinnecker H. (1999) \aj, {\it 117}, 2381-2397.

\refs Preibisch Th., Guenther E., Zinnecker H., Sterzik M., Frink S., and R\"oser S. (1998) \aap, {\it 333}, 619-628.

\refs Preibisch Th., Brown A.~G.~A., Bridges T., Guenther E., and Zinnecker H. (2002) \aj, {\it 124}, 404-416.

\refs Preibisch Th., Kim Y-C., Favata F., Feigelson E.~D., Flaccomio E. et al. (2005) \apjs, {\it 160}, 401-422.

\refs Podosek F.~A. and Cassen P. (1994) {\it Meteoritics, 29}, 6-25.

\refs Rebull L.~M., Hillenbrand L.~A., Strom S.~E., Duncan D.~K., and Patten B.~M. et al. (2000) \aj, {\it 119}, 3026-3043.

\refs Scalo J.~M. (1998) In {\it The stellar intial mass function}
(Gilmore G. and Howel D., eds.), {\it 142}, pp. 201-236. ASP, San Francisco.

\refs Scoville N.~Z. and Hersh K. (1979) \apj, {\it 229}, 578-582.


\refs Sanduleak N. (1971) \pasp, {\it 83}, 95-97.

\refs Selman F.~J. and Melnick J. (2005) \aap, {\it 443}, 851-861.

\refs Sherry W.~H. (2003) {\it PhD Thesis}, SUNY, Stony Brook.

\refs Sherry W.~H., Walter F.~M., and Wolk S.~J. (2004) \aj, {\it 128}, 2316-2330.

\refs Shu F.~H. and  Lizano S. (1988) {\it Astrophys. Letters Commun., 26}, 217-226.

\refs Shu F.~H., Adams F.~C., and Lizano S. (1987) \araa, {\it 25}, 23-81.

\refs Sicilia-Aguilar A., Hartmann L.~W., Brice\~no C., Muzerolle J., and Calvet N. (2004) \aj, {\it 128}, 805-821.

\refs Sicilia-Aguilar A., Hartmann L.~W., Hern\'andez J., Brice\~no C., and Calvet N. (2005) \aj, {\it 130}, 188-209.

\refs Sicilia-Aguilar A., Hartmann L.~W., Calvet N., Megeath S.~T., Muzerolle J., et al.  (2006) \aj, in press.

\refs Siess L., Dufour E., and Forestini M. (2000) \aap, {\it 358}, 593-599.

\refs Sivan J.~P. (1974) {\it Astron. Astrophys. Suppl. Ser., 16}, 163-172.

\refs Slawson R.W. and Landstreet J~.D. (1992) {\it Bull. Am. Astron. Soc., 24}, 824-824.

\refs Slesnick C.~L., Carpenter J.~M., and Hillenbrand L.~A. (2005) In 
{\it PPV Poster Proceedings} \\
http://www.lpi.usra.edu/meetings/ppv2005/pdf/8365.pdf

\refs Smith L.~J. and Gallagher J.~S. (2001) \mnras, {\it 326}, 1027-1040.

\refs Soderblom D.~R. (1996) In {\it Cool stars, Stellar Systems and the Sun IX}
(R. Pallavicini and A.~K. Dupree, eds.), {\it 109}, pp. 315-324. ASP, San Francisco.

\refs Soderblom D.~R., Jones B.~F., Balachandran S., Stauffer J.~R., Duncan D.~K. et al. (1993) \aj, {\it 106}, 1059-1079.

\refs Stolte A., Brandner W., Grebel E.~K., Lenzen R., and Lagrange A.~M. (2005) \apj, {\it 628}, L113-L117.

\refs Strom S.~E., Strom K.~M., and Grasdalen G.~L. (1975) \araa, {\it 13}, 187-216.

\refs Strom S.~E., Edwards S., and Skrutskie M.~F. (1993) In {\it Protostars and Planets III}
(E. H. Levy and J. I. Lunine, eds.), pp. 837-866. Univ. of Arizona, Tucson.

\refs Uchida K.~I., Calvet N., Hartmann L., Kemper F., Forrest W.~J., et al.
 (2004) \apjs, {\it 154}, 439-442.

\refs Vanhala H.~A.~T. and Cameron A.~G.~W. (1998) \apj, {\it 508}, 291-307.


\refs Walborn N.~R., Barb\'a, R.~H., Brandner W., Rubio M., Grebel E.~K., and Probst R.~G. (1999) \aj, {\it 117} 225-237.

\refs Walter F.~M. and Kuhi L.~V. (1981) \apj, {\it 250}, 254-261.

\refs Walter F.~M. and Myers P.~C. (1986) In {\it IV Cambridge Workshop 
on Cool Stars, Stellar Systems, and the Sun}(M. Zeilik and D.~M. Gibson, eds.), {\it 254}, pp.55-57.
Springer-Verlag, Berlin-Heidelberg-New York.

\refs Walter F.~M., Brown A., Mathieu R.~D., Myers P.~C., and Vrba F.~J. (1988) \aj, {\it 96}, 297-325.

\refs Weaver W.~B. and Babcock A. (2004) \pasp, {\it 116}, 1035-1038.


\refs White R.~J. and Basri G. (2003) \apj, {\it 582}, 1109-1122.

\refs White R.~J. and  Hillenbrand L.~A. (2005) \apj, {\it 621}, L65-L68.

\refs Wilking B.~A., Schwartz R.~D., and Blackwell J.~H. (1987) \aj, {\it 94}, 106-110.

\refs Wiramihardja S.~D., Kogure T., Yoshida S., Ogura K., and Nakano M. (1989) \pasj, {\it 41}, 155-174.

\refs Wiramihardja S.~D., Kogure T., Yoshida S., Ogura, K., and Nakano M. (1993) \pasj, {\it 45}, 643-653.

\refs Yamaguchi R., Mizuno N., Onishi T., Mizuno A., and Fukui, Y. (2001) \pasj, {\it 53}, 959-969.

\refs Zinnecker H., McCaughrean M.J., and Wilking B.A. (1993) In
{\it Protostars and Planets III} (E.H.~Levy and J.I.~Lunine, eds.), pp. 429-495.
Univ. of Arizona, Tucson.

\end{document}